%% file: 89Her_circum.tex
\documentclass[fleqn,usenatbib]{mnras}

\usepackage{mathptmx}
\usepackage{pdflscape}

\usepackage[T1]{fontenc}
\usepackage{ae,aecompl}

\usepackage{graphicx}	
\usepackage{amsmath}	
\usepackage{amssymb}	

\title[]{Long-term photospheric instabilities and envelopes dynamics in the post-AGB binary system 89 Herculis.}

\author[M. Gangi et al.]{
M. Gangi,$^{1}$\thanks{E-mail: manuele.gangi@inaf.it}
M. Giarrusso,$^{2}$
M. Munari,$^{3}$
C. Ferrara,$^{3,4}$
C. Scalia,$^{3,4}$ 
F. Leone$^{3,4}$
\\
$^{1}$INAF - Osservatorio Astronomico di Roma, Via Frascati 33, I-00078 Monte Porzio Catone, Italy \\
$^{2}$INFN - Laboratori Nazionali del Sud, Via S. Sofia 62, I-95123 Catania, Italy \\
$^{3}$INAF - Osservatorio Astrofisico di Catania, Via S. Sofia 78, I--95123 Catania, Italy \\
$^{4}$Universit\`a di Catania, Dipartimento di Fisica e Astronomia, Sezione Astrofisica, Via S. Sofia 78, I--95123 Catania, Italy
}

\date{Accepted XXX. Received YYY; in original form ZZZ}

\pubyear{2015}
\begin{document}
\label{firstpage}
\pagerange{\pageref{firstpage}--\pageref{lastpage}}
\maketitle

\begin{abstract}
We present a long-term optical spectroscopic study of the post-AGB binary system 89 Herculis, with the aim to characterize the relationship between photospheric instabilities and dynamics in the close circumstellar environment of the system. This study is based on spectra acquired with the high-resolution Catania Astrophysical Observatory Spectropolarimeter and archive data, covering a time interval between 1978 and 2018. We find long-term changes in the radial velocity curve of the system, occurring mostly in amplitude, which correlate with the variability observed in the blue-shifted absorption component of the P Cygni like H$\alpha$ profile. Two possible scenarios are discussed. We also find strong splitting in the s-process elements of \ion{Ba}{ii} $6141.713$ \AA\ and $6496.898$ \AA\,lines, with short-term morphological variations. A Gaussian decomposition of such profiles allows us to distinguish four shell components, two expanding and two in-falling toward the central star, which are subject to the orbital motion of the system and are not affected by the long-term instabilities. Finally, we find that the numerous metal lines in emission could originate in regions of a structured circumbinary disk that have sizes proportional to the energy of the corresponding upper level transition $\rm E_{up}$. This study demonstrates the potential of long-term high-resolution spectroscopy in linking together the instability processes occurring during the late evolutionary stages of post-AGBs and the subsequent phase of PNe.
\end{abstract}




\section{Introduction}
Stars with a main-sequence initial mass in the $0.8-8\,\rm M_{\odot}$ range evolve from the Asymptotic Giant Branch (AGB) to the Planetary Nebulae (PNe) stages. The end of the AGB is characterized by a phase of very strong mass loss ($10^{-7}-10^{-4}\,\rm M_{\odot}\,yr^{-1}$) that occurs in the form of gas and dusty wind driven by stellar radiation pressure \citep[see reviews by][]{Winckel2018,Hofner2018}. This process seems to be neither isotropic nor steady over time and important episodic mass-loss enhancements can occur. The subsequent phase of post-AGB is then characterized by a spatially complex environment, that may include long-lived quasi-Keplerian stratified disks, short-term structures, equatorial overdensities and axisymmetric bipolar jets. 

To explain the rich variety of morphologies of PNe, a wide range of dynamical processes have been invoked as shaping agents \citep{Balick2002}. Among these, the attractive possibility of large scale magnetic fields \citep{Jordan2005} on the surface of the central star has been ruled out with systematic spectropolarimetric campaigns \citep{leone2011,Leone2014,Ramos2014,gonzalez2015}. On the contrary, binary interaction has emerged as the more fundamental process \citep[e.g.][]{DeMarco2009}. Unseen companions are indeed suspected to play a key role in the outflow processes, also influencing the time scales of the different evolutionary stages of the primary star \citep{Olivier2013}. Binarity seems also to affect the dynamics of disks as it has been highlighted in the particular case of binary post AGBs \citep[][and references therein]{Kluska2019}; disks are found to be, in turn, mainly responsible for the orbital evolution of the binary system \citep[e.g.,][]{Dermine2013}.

Mass loss is also a key ingredient for the understanding of the chemical abundance in evolved stars. Indeed, many post-AGB stars are found to be under-abundant of s-process elements, that are incorporated into refractory grains. Such a depletion can then mask the enrichments of the nucleosynthetic products formed in the previous AGB state \citep{Winckel2018}. Even if the involved complex processes are still a matter of discussion, it seems necessary that there exist a stable disk around the post-AGB star, at least partly, responsible for the observed chemical anomaly \citep{Gezer2015}. Indeed, the radiation pressure acting on the circumstellar dust grains can clean the dust from the gas. The latter, in turn, can be re-accreted into the stellar surface \citep{Winckel2018}. 

From an observational point of view, a wealth of high spatial resolution techniques (e.g. high-contrast imager and interferometry in the optical, infrared, millimetric and radio wavelengths, thanks to new advanced adaptive optics and post-processing techniques) can provide important constrains on models. However, the achievable contrast and spatial resolution are not yet sufficient to be able to investigate the environment in proximity (i.e. few au) of the central star and their mutual relationship.

An important step forward this kind of studies can be achieved by looking at optical/infrared high resolution spectroscopy. Indeed, a variety of peculiarities have been taken as direct evidence of circumstellar envelope and mass loss manifestations: i) variable and complex profiles of Balmer and sodium D lines, which include absorption and emission components, ii) electron transitions of low excitation potential iron-peak and s-process elements result in largely asymmetric spectral line profiles with several components, iii) forbidden metal emission lines \citep{Klochkova2014}. In addition, a variability in the aforementioned features could be associated with the kinematics of the circumstellar envelope and its morphology. This aspect however requires high-quality data that span over wide temporal bases.

The goal of this paper is to investigate the relationship between spectroscopic variability and dynamics in the circumstellar environment of the most known post-AGB binary system: 89 Herculis (Section \ref{sec:89Herculis}). This study is based on high resolution optical spectroscopy, with data spanning a time interval of more than 25 years (Section \ref{sec:DATA}). In particular, we first focus on the long-term variability, in terms of radial velocities (Section \ref{sec:RadVel}) and H$\alpha$ line morphological changes (Section \ref{sec:HALPHA}). Then, we analyze the short-term variability in the morphologies of s-process elements lines (Section \ref{sec:Selem}) and in the kinematics of metal lines in emission (Section \ref{sec:Emission}). Finally, we discuss the results in Section \ref{sec:Discussion}, giving our conclusion in Section \ref{sec:Conclusion}.

\input{Tables/RadialVelocities.tex}

\section{89 Herculis}\label{sec:89Herculis}
The F2Ibp star 89 Herculis (HR 6685, HD 163506) is considered the prototype of the new class of post-AGB binaries surrounded by a circumbinary dust disk \citep{Waters1993}. The primary component pulsates with a period of $63.5\, \rm d$ \citep{Ferro1983}, while the orbital period of the system is $289.1\, \rm d$ days \citep{Oomen2018}. The mass-loss rate derived by  \citet{Sargent1969} from the P Cygni type H$\alpha$ profile is $\sim 10^{-8}$ $\rm M_{\sun}$ $\rm yr^{-1}$. The primary has a mass of $\rm M_p = 0.6$ $\rm M_{\sun}$ and a radius of $\rm R_p = 71$ $\rm R_{\sun}$, while the secondary is supposed to be a main-sequence star with a mass of $\rm M_s = 0.4$ $\rm M_{\sun}$ \citep{Hillen2014}.

The structure of the circumbinary envelope was investigated interferometrically by \citet{Bujarrabal2007} and \citet{Hillen2013,Hillen2014}. Both research teams suggested the presence of at least two nebular components: an expanding hour-glass structure and an unresolved circumbinary Keplerian disk. The hour-glass axis is tilted with respect to the Line-of-Sight (LoS) of $\rm \theta_{LoS} \sim 12^{\circ}$ with a Position Angle $\rm \theta_{PA}=45^{\circ}$.

\section{Observational data}\label{sec:DATA}
From 2014 to 2018, we have performed optical high resolution spectroscopy of 89 Herculis with the high resolution \emph{Catania Astrophysical Observatory Spectropolarimeter} \citep[CAOS, $\rm R = 55 000$,][]{Leone2016a} at the 0.91 m telescope of the stellar station of the Catania Astrophysical Observatory (\emph{G. M. Fracastoro} Stellar Station, Serra La Nave, Mt. Etna, Italy). Further spectra were also acquired on June, $\rm 2^{th}$, 2012, with the Device Optimized for the LOw RESolution spectrograph \citep[DOLORES, LRS in short, $\rm R=4300$,][]{Molinari1997}, on May $\rm 27^{th}$, 2017 with the HArps-North POlarimeter \citep[HANPO, $\rm R=115000$,][]{Leone2016b} of the Telescopio Nazionale Galileo (Roque de Los Muchachos Astronomical Observatory, La Palma, Spain) and on June, $\rm 25^{th}$, 2018, with the High Accuracy Radial Velocity Planet Searcher spectrograph \citep[HARPS-N, $\rm R=120000$,][]{Cosentino2012}.

Data were reduced according to the standard procedures using the \textsc{noao/iraf}\footnote{IRAF is distributed by the National Optical Astronomy Observatory, which is operated by the Association of the Universities for Research in Astronomy, inc. (AURA) under cooperative agreement with the National Science Foundation.} packages. Each spectrum was previously corrected for the heliocentric velocity due to the Earth's motion by means o the \textsc{iraf} task \emph{rvcorrect}.

In addition, reduced data of 89 Her have been retrieved from all available archives. In particular from the
\begin{itemize}
\item Ritter Observatory Archive\footnote{http://astro1.panet.utoledo.edu/$\sim$wwritter/archive/PREST-archive.html}. Spectra were collected from 1993 to 1995 with the echelle spectrograph of the 1m Ritter Observatory Telescope ($\rm R=26000$);
\item BTA\footnote{https://www.sao.ru/oasis/cgi-bin/fetch?lang=en} (Big Telescope Alt-azimuth) archive. Spectra have been acquired in 1998 and 2010 with the Nasmyth Echelle Spectrograph \citep[NES, $\rm R=60000$,][]{Panchuk2007};
\item ELODIE\footnote{http://atlas.obs-hp.fr/elodie/}  archive. A spectrum was acquired on August 18, 2004 with the ELODIE spectrograph at the Observatoire de Haute-Provence 1.93m Telescope \citep[$\rm R=42000$,][]{Moultaka2004};
\item Canadian Astronomy Data Centre. Spectra were collected from 2005 to 2009 at the 3.6 m Canadian-France-Hawaii Telescope with the "Echelle SpectroPolarimetric Device for the Observation of Stars" \citep[ESPaDOnS, $\rm R=68000$,][]{Donati2006};
\item ESO\footnote{http://archive.eso.org/eso/eso\_archive\_main.html} archive. A spectrum was acquired on May 9, 2013 at the MPG/ESO 2.2m Telescope with the "Fiber-fed Extended Range Optical Spectrograph" \citep[FEROS, $R=48000$,][]{Kaufer1999}.
\end{itemize}

The complete logbook of the observations is given in Table \ref{tab:RadVel}.

\section{Long-term Radial velocities variability}\label{sec:RadVel}
Since the 1920s the radial velocity (RV) variability of 89 Herculis has been investigated \citep{Burki1980, Fernie1981, Ferro1984, Waters1993}. From these works a periodicity of about 288 days, ascribable to the binary nature of the system, was established. Using data acquired with \textsc{coravel} in the period 1978-1992, \citet{Waters1993} has determined the orbital parameters of the system. Recently, \citet{Oomen2018} have revised the orbital parameters by acquiring new data with \textsc{hermes} in the period 2009-2017. However, important scatter around the mean RV curve (see, for example, figure 1 of \citealt{Waters1993} or figure B.1 of \citealt{Oomen2018}) has always been found and has been attributed to pulsations or irregular instabilities in the extended atmosphere of the star. To overcome the problem of the irregular scatter of the data and thus to better constrain the orbital parameters, measurements over many orbital cycles are necessary. For this reason, thanks to the relatively large temporal base of our data, we can perform a revision of the orbital parameters through the study of the RV variability of 89 Her.

Following \citet{Catanzaro2019}, we measured radial velocities of 89 Her by cross-correlating each observed spectrum with a synthetic template. The latter was computed by means of the \textsc{synthe} \citep{Kurucz1993} code, using an atmospheric model with effective temperature $\rm T_{eff}=6500$  $\rm K$, surface gravity $\rm \log g=1.0$ and iron abundance $\rm [Fe/H]=-0.5$, as measured by \citet{Kipper2011}. Cross-correlations were performed with the \textsc{iraf} task \emph{fxcor} in the widest spectral range available, excluding regions containing Balmer and telluric lines as well as transitions that are strongly influenced by envelope components (e.g. the Na D-lines at $\rm 589$ and $\rm 590$ $\rm nm$ or lines of s-process elements). Measured radial velocities are reported in Table \ref{tab:RadVel}. In this study, we have also included radial velocities by \citet{Burki1980} and \citet{Waters1993} to span about 40 years. 

\begin{figure}
\begin{center}
\includegraphics[trim=0cm 0.3cm 7.2cm 18.2cm, clip=true,width=8cm,angle=0]{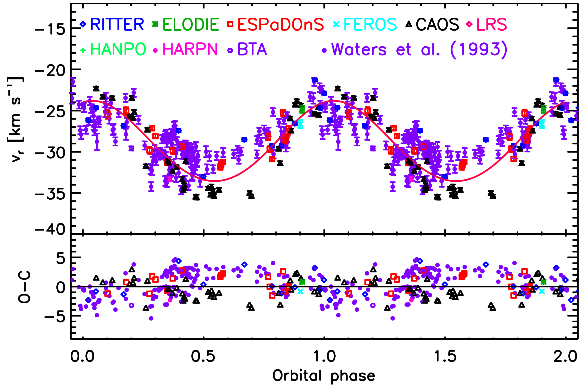}
\begin{center}\caption{\label{fig:VelALL} Phase-folded radial velocities of 89 Herculis and theoretical mean curve. The residuals are shown in the bottom panel.}\end{center}
\end{center}
\end{figure}

\begin{figure}
\begin{center}
\includegraphics[trim=1.6cm 0.9cm 8.2cm 5.5cm, clip=true,width=6cm,angle=0]{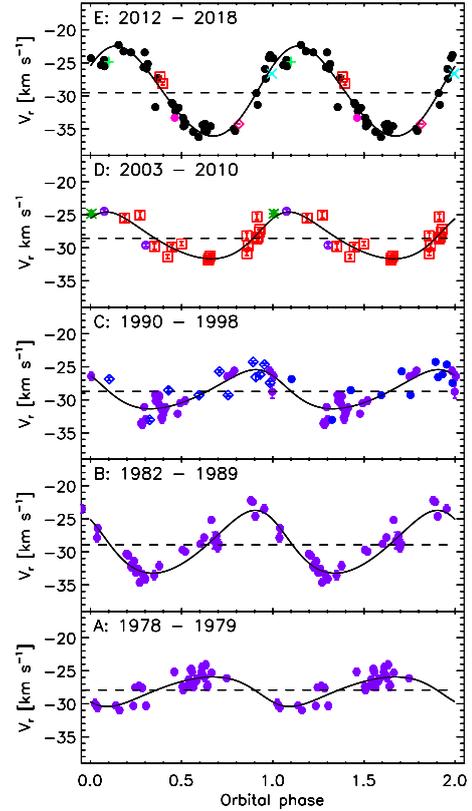}
\begin{center}\caption{\label{fig:RadVel} Phase-folded radial velocities for different epochs of 89 Herculis. Colors and symbols are as in Fig. \ref{fig:VelALL}. Theoretical curves, computed with the orbital parameters reported in Table \ref{tab:OrbitalParameters}, are superimposed.}\end{center}
\end{center}
\end{figure}

From a general look at the data, we confirm the presence of significant scatter around the mean RV curve, which can be more than $5\, \rm km\, s^{-1}$ (Fig. \ref{fig:VelALL}). However we found out that this spread is not randomly distributed over the years. Indeed, there are time intervals where the amplitude of the velocity curve increases. In particular, by dividing the data in the epochs $\rm A=[1978,1979]$, $\rm B=[1982,1989]$, $\rm C=[1990,1998]$, $\rm D=[2003,2010]$ and $\rm E=[2012,2018]$, the amplitudes in B and E are about twice those observed in the other epochs (Fig. \ref{fig:RadVel}). 

\subsection{Orbital solutions}

Following \citet{Hilditch2001}, we determine the orbital parameters for the binary system (i.e. the epoch of the periastron passage $\rm T$, the orbital period $\rm P$, the systemic velocity $\rm \gamma$, the velocity semi-amplitude $\rm K$, the eccentricity of the orbit $\rm e$ and the longitude of periastron $\rm \omega$) from the RV curve through the following equations:
\begin{eqnarray}
\rm V_{rad} & = & \rm \gamma + \rm K[\cos{(\rm \theta + \rm \omega)} + \rm e \cos{\omega}]\label{eq:OrbSol1} \label{eq:RadVel}\\
\rm K & = & \frac{\rm 2 \rm{\pi} \rm a \sin{\rm i}}{\rm P\sqrt{1-\rm e^2}},\label{eq:OrbSol2}
\end{eqnarray}
with $\rm \theta$ the true anomaly, $\rm a$ the semi-major axis of the orbit and $\rm i$ the orbital inclination. From $\rm P$ and $\rm K$ it is also possible to determine the binary mass function $\rm f$:
\begin{eqnarray}
\rm f & = \frac{\rm M_2^3 \sin^3{\rm i}}{(\rm M_1+\rm M_2)^2}  = \frac{\rm P \rm K^3}{\rm 2 \pi \rm G}\label{eq:OrbSol3}
\end{eqnarray}

We have made an independent fit of Eq. \ref{eq:OrbSol1} by considering the RV curves divided into the aforementioned epochs. According to \citet{Catanzaro2018}, fits have been performed using the amoeba method \citep{Nelder1965} while errors were statistically estimated as the variation in the parameters that increases by 1 the $\rm \chi^2$ value.

In Fig. \ref{fig:RadVel} we report the RV curves with the theoretical solutions, while the corresponding orbital parameters are reported in Table \ref{tab:OrbitalParameters}. Scatters around the mean RV curve are now reduced within about $3\, \rm km\, s^{-1}$. 

The important differences among the obtained RV curves can be interpreted as a consequence of the different data sampling and number of RV points within the A-E epochs. Moreover, in the A and C epochs the available data do not completely cover all the orbital phases. As a consequence, pulsations or irregular scatters can dominate over the global shape of the fitted curve, leading to a wrong determination and interpretation of orbital parameters changes. On the other hand we stress that the residuals of the cumulative fit of the measurements (see Fig. \ref{fig:VelALL}, bottom), are not randomly distributed and that the increase in amplitude in the epochs B and E can hardly be interpreted as a sampling problem. Rather, we think that it represents a real change in the amplitude. 

In light of this, in Fig. \ref{fig:FitOrbPar} we report the fitted parameters against time, considering the mean year for each of the epochs A-E. From a general look, the eccentricity $\rm e$ and the period $\rm P$ do not show any appreciable variability. On the contrary, the systemic velocity $\rm \gamma$ and the semi-amplitude $\rm K$ present a significantly increase of their values from the D to E epoch. Finally, the epoch of the periastron passage $\rm T$ and the longitude of periastron $\rm \omega$ show a constant decrease over time. In the following we discuss the possible scenarios. 

\subsection{The Nature of the RV modulation}\label{sec:RVmodulation}
The observed variations can be explained in several ways. The possible hypotheses for these are: (i) stellar pulsations (ii) the presence of a third body or (iii) an intrinsic variation of the orbital parameters of the binary system. In the following we will discuss these three scenarios.

\subsubsection{Pulsations}
It has been shown that a long or short-term RV modulation can be a consequence of atmospheric velocity gradient perturbations in pulsating stars. Indeed, modulated variability was discovered both in classical and in long-period Cepheids \citep[e.g.][and references therein]{Derekas2012,Anderson2016}, as well as in other types of pulsating stars, such as $\delta$ Sct or $\gamma$ Dor \citep[e.g.][]{Bowman2016, Guzik2016}.

In principle, due to the pulsating nature of 89 Herculis, the observed RV modulation should be ascribed to the possible presence of long-term cycle-to-cycle pulsation variations. In such a case the atmospheric velocity perturbations should lead to important changes in the morphology of the spectral lines during the different epochs. To investigate this possibility we have searched for a variation in the photospheric absorption spectral lines by calculating the equivalent widths ($\rm EWs$) and the bisector inverse span ($\rm BIS$) in the epochs $\rm D$ and $\rm E$, for different absorption lines with different depths. The BIS were computed according to \citet{Anderson2016}, i.e. as the difference between bisector velocities at the top 10-40 and bottom 60-90 $\%$ normalized flux of the absorption profile.

Fig. \ref{fig:BIS_4508} shows an example of EWs and BIS calculated for the strong line of \ion{Fe}{ii} at $4508.2805$ \AA; similar results are for absorption lines of other neutral and ionised species. As expected for pulsating stars, we have found that BIS is variable with the pulsation period $\rm P_{puls}$. In particular we have found $\rm P_{puls} = 63.61 \pm 0.01\, \rm d$, in full agreement with the photometric pulsation period found by \citet{Waters1993}. On the contrary, EWs of the same lines do not show any appreciable variability, neither pulsational nor orbital. In both cases, however, we have found no significant difference in the behaviour of BIS and EWs between epochs $\rm D$ and $\rm E$; the slightly lower values in EWs of the period $\rm D$ are ascribable to the continuum determination in ESPaDOnS and CAOS spectra because of the different spectral resolving power.

We conclude then that pulsations are unlikely to be the major contributors to the observed RV modulation.

\begin{figure}
\begin{center}
\includegraphics[trim=0.7cm 2.cm 10.5cm 6.7cm, clip=true,width=7cm,angle=0]{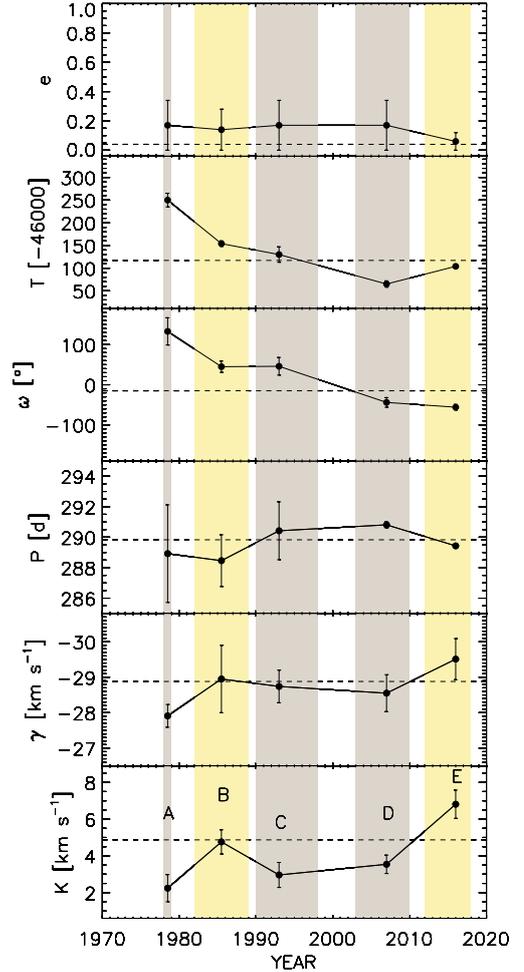}
\begin{center}\caption{\label{fig:FitOrbPar} Variation of Orbital parameters over the years. Horizontal dashed lines represent values found through the cumulative fit of the data (Fig. \ref{fig:VelALL}). Colored regions indicate time intervals where RV curves are available (yellow: high amplitude RV curves, gray: low amplitude RV curves. See Fig. \ref{fig:RadVel}).}\end{center}
\end{center}
\end{figure}

\begin{figure}
\begin{center}
\includegraphics[trim=14cm 5cm 0.5cm 3.6cm, clip=true,width=6cm,angle=180]{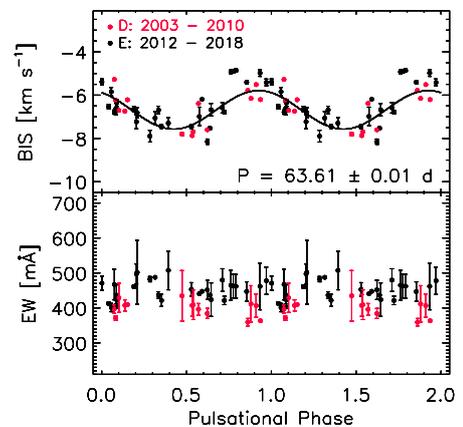}
\begin{center}\caption{\label{fig:BIS_4508} BISector (top) and Equivalent Widths (bottom) of the \ion{Fe}{ii} line at $4508.2805$ \AA.}\end{center}
\end{center}
\end{figure}

\begin{figure}
\begin{center}
\includegraphics[trim=0.7cm 8.cm 10.5cm 12.7cm, clip=true,width=7cm,angle=0]{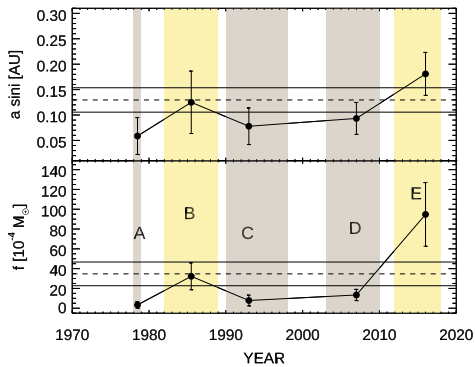}
\begin{center}\caption{\label{fig:FitMassFunc} Variability of $\rm a \sin{i}$ (top) and binary mass function (bottom).  Horizontal dashed lines represent values from a cumulative fit of the data (Fig. \ref{fig:VelALL}). Colored regions indicate time intervals where RV curves are available (yellow: high amplitude RV curves, gray: low amplitude RV curves. See Fig. \ref{fig:RadVel}). }\end{center}
\end{center}
\end{figure}

\input{Tables/OrbitalParameters_NEW_VERSION.tex}

\subsubsection{Presence of a third body}
In principle, the presence of a third body in a hierarchical triple star system can be responsible of RV modulation. In such a case the constant value of $\rm \gamma$ in Eq. \ref{eq:RadVel} should be considered as a variable velocity with the same analytic form of Eq. \ref{eq:RadVel}. In this case $\rm \gamma$ represents the center-of-mass velocity of the close binary system orbiting the third body and it is expected to be variable according to the orbital period of the third component.

In our case, $\rm \gamma$ has a constant value in the epochs B, C and D with a decreasing and an increasing of only about 1 $\rm km$ $\rm s^{-1}$ in the A and E epochs, respectively (see Table \ref{tab:OrbitalParameters} and Fig. \ref{fig:FitOrbPar}). While such a behavior may still be compatible within the picture of a hierarchical triple star system, we exclude the possibility that  during the time interval in which these small velocity changes occur the presence of a third body represents the major contributor to the observed variation in the RV curves.

\subsubsection{Intrinsic variations in the orbital solutions}
As discussed above, $\rm T$ and $\omega$ are the only parameters that show a constant decreasing over time, while $\gamma$ and $\rm k$ show an increase only in the A-B and D-E epochs (Fig. \ref{fig:FitOrbPar}).

The variation of $\rm T$ and $\omega$ suggests that the orbit precedes in the prograde sense. In particular our data indicate that a variation of $\Delta \omega = 189^\circ$ is made in $\sim 40$ years, suggesting that a complete rotation can be made in $\sim 76$ years. The orbital plane rotation, however, has no effect on the other parameters.

From equations \ref{eq:OrbSol2} and \ref{eq:OrbSol3} we have computed the binary mass function and the $\rm a \sin{i}$; values are reported in Table \ref{tab:OrbitalParameters}. Their variations are shown in Fig. \ref{fig:FitMassFunc} and follow the behaviour of $\rm K$, a direct consequence of Eqs. \ref{eq:OrbSol2} and \ref{eq:OrbSol3}, since $\rm P$ and $\rm e$ can be considered constant. Two different scenarios could explain the observed trend: (i) a variation of the inclination of the binary orbit or (ii) a variation of the masses.

In the first case, assuming an average value of $\rm i = 12^\circ$, the semi-major axis of the orbit would be about $\rm 0.5$ au. With this value, we found that during the A-B and the D-E epochs the orbit plane inclination would increase from about $\rm 6^\circ$ to $\rm 14^\circ$ and from about $\rm 10^\circ$ to $\rm 20^\circ$, respectively. In such case, the observed RV amplitude modulation would be then related to an orbital plane wobbling of $\sim \rm 10^\circ$ in a time interval of $\sim 13$ yrs.

Alternatively, with $\rm i=const$, the important increase of the binary mass function during epochs A-B and D-E would be a consequence of mass transfer between the components of the system. Such mass transfer could be triggered from the mass-loss of the primary star. In particular, if we consider $\rm M_1+M_2=const$, and assume $\rm M_1=0.6$ $\rm M_{\sun}$ for the primary and $\rm M_2=0.4$ $\rm M_{\sun}$ for the secondary \citep{Hillen2014}, this would mean that during the A-B and the D-E epochs the secondary must undergone an increase of mass of the order of $\rm 10^{-1} M_2$, which corresponds to a mass loss rate of the primary of $10^{-2}$ $\rm M_{\sun}$ $\rm yr^{-1}$. Such a value, however, is not compatible with the typical mass-loss rates expected for post-AGB objects.

We conclude that the observed RV modulation could be explained with an orbital plane wobbling while it can hardly be correlated with a mass transfer between the components of the system. The behaviour of the $\rm T$ and $\rm \omega$ parameters only supports the hypothesis of an orbital plane rotation, a scenario that should be further investigated by appropriate models. The cumulative fit of the RV data (Fig. \ref{fig:VelALL}) mitigates long-term and sporadic variation effects, and it should be taken as the best measurement of the orbital parameters of 89 Her.

To further investigate the nature of the RV modulation, we explore now the possibility that the already known long-term spectroscopic variability of the system can be associated with these changes in RV amplitude. 

\section{Long-term H$\alpha$ variability}\label{sec:HALPHA}
Several studies have been carried out on 89 Herculis spectral variability. 
In particular \citet{Bohm1956} firstly pointed out important variations in the radial velocity amplitudes and in the structure of the Balmer and sodium D lines in the period 1951-1955. The author had also speculated on a possible correlation between these two evidences. Variations in the structure of Balmer and sodium D lines were later ascribed to optically thick expanding shells and then interpretable as episodic mass loss enhancements from the primary star \citep{Sargent1969}. \citet{Khalilov2010} discovered a long-period variability in EWs of the H$\alpha$ P Cygni-like profile of 89 Her, finding a period of $6553.6\, \rm d$ for the absorption component and a period of $10922.7\, \rm d$ for the emission one. The EWs of the Na I D absorption line were also found to be variable with a period of about $5000\, \rm d$ \citep{Khalilov2019}.

To verify the hypothesis made by \citet{Bohm1956} on a possible relationship between the variability of the RV curves and the episodic mass loss enhancements of the star, we now analyze the long-term H$\alpha$ line profile changes.

\input{Tables/Logbook_EW_ABS_EM.tex}

The H$\alpha$ P Cygni-like profile is considered the most suitable diagnostic for wind and mass-loss rate studies. Indeed such a profile indicates the presence of a strong stellar wind which can produce the emission line and the blue-shifted absorption. The latter is produced by optically thick expanding shells, along the line of sight, and it is then linked to the flow velocities as well the mass-loss rate of the gas component. Assuming a spherical expansion with constant velocity, the mass-loss rate $\rm \dot M$ can be related to the properties of the blue-shifted absorption component by the continuity equation \citep{Hofner2018}:
\begin{eqnarray}
\rm \dot M = 4 \pi r^2 v_{\infty} \rho(r) \label{eq:continuity}
\end{eqnarray}

where $\rm v_{\infty}$ and $\rho$ are the terminal velocity and the density of the gas component, respectively. Terminal velocity can be derived from the position of the blue absorption edge of the H$\alpha$ profile, while density is related to the depth of the absorption component; the product $v_{\infty} \rho$ can be then estimated from the measurement of the equivalent width of the absorption component.

\begin{figure}
\begin{center}
\includegraphics[trim=13.5cm 0.5cm 1.5cm 7.8cm, clip=true,width=6cm,angle=180]{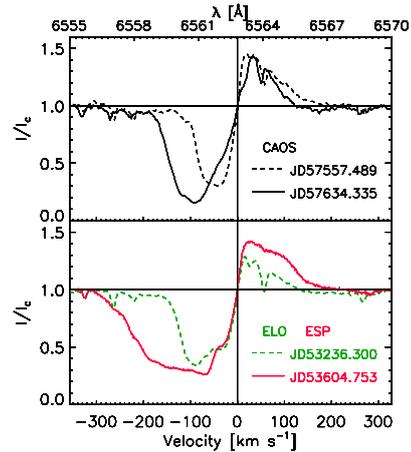}
\caption{\label{fig:Halpha} Examples of H$\alpha$ profiles variations in a time interval of 77 days (top) and 368 days (bottom).}
\end{center}
\end{figure}

The variations across the H$\alpha$ P Cygni profile of 89 Her cover different time scales, from days to years, with remarkable changes in the intensity and in the structure of the blue-displaced component. Fig. \ref{fig:Halpha} shows examples of such variations. We perform a crude estimate of the equivalent width of the absorption ($\rm EW_{abs}$) and emission ($\rm EW_{em}$) components. They were measured in the fixed intervals from $-400$ to $0$ $\rm km\, s^{-1}$ and from $0$ to $245$ $\rm km\, s^{-1}$, respectively. The emission components have been firstly corrected for the presence of $\rm H_2O$ telluric lines at $6563.521$ \AA, $6564.061$ \AA, $6564.203$ \AA. HJD 57666.248, 58251.592 and 58268.548 data were excluded since the profiles were significantly altered. We also measured the terminal velocity $\rm v_{\infty}$ as the one corresponding to the blue wing of the component at a flux of $20 \%$ less than the normalized continuum. Measurements are reported in Table \ref{tab:ew}.

\begin{figure}
\begin{center}
\includegraphics[trim=13.cm 2.1cm 1.5cm 4cm, clip=true,width=8cm,angle=180]{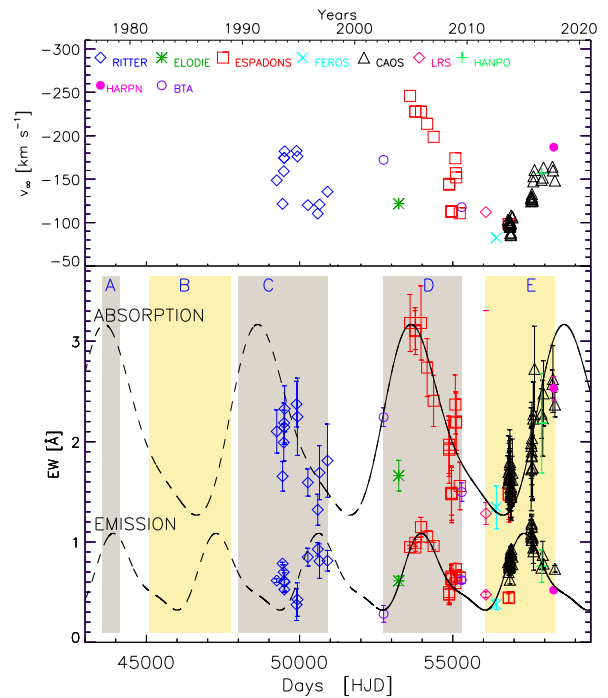}
\begin{center}\caption{\label{fig:EW} Top: terminal velocity of the H$\alpha$ absorption component. Bottom: Equivalent Widths of the H$\alpha$ absorption and emission components against time. Colored regions indicate time intervals where RV curves are available (yellow: high amplitude RV curves, gray: low amplitude RV curves. See Fig. \ref{fig:RadVel}).}\end{center}
\end{center}
\end{figure}

Fig. \ref{fig:EW} (bottom) shows the $\rm EWs$ in time. Considering data acquired in the last 20 years (regions D and E), two episodes of important variations in the $\rm EWs$ are shown. The variations are mostly pronounced in the $\rm EWs$ of the absorption component, with an amplitude that is approximately double than that of the emission one. Double-sine curves are drawn to better guide the eye. The $\rm EWs$ of the absorption component exhibits a peak-to-peak variation in a time interval of about $5000\, \rm d$ while the emission component as a variation in $3300\, \rm d$. An extrapolations of the curves back in time to include \textsc{ritter} data (region C), does not allow us to conclude that these variations are strictly periodic and further data spread over a very long timebase are necessary to clarify this possibility. We cannot then confirm the results found by \citet{Khalilov2010}, i.e. a periodicity of 6553.3 d and 10922.7 d for the absorption and emission component, respectively. It is intriguing to note, however, that the period we found for the H$\alpha$ $\rm EW_{abs}$ variability is the same as that found by \citet{Khalilov2019} for the absorption components of the sodium D lines.

Fig. \ref{fig:EW} (top) shows the terminal velocity $\rm v_{\infty}$ of the blue-displacement absorption component against time. The variations follow the behaviour of the corresponding $\rm EW_{\rm abs}$, with values between $-80$ $\rm km$ $\rm s^{-1}$ and $-250$ $\rm km$ $\rm s^{-1}$. 

From the observed $\rm EWs_{\rm abs}$ variation we can now estimate the variation in the mass-loss rate. Assuming the validity of Eq. \ref{eq:continuity} and considering the curve shown in Fig. \ref{fig:EW}, we can express the ratio between the maximum and the minimum mass-loss rate as:

\begin{eqnarray}
\frac{\rm \dot M_{max}}{\rm \dot M_{min}} = \frac{\rm EW_{max}}{\rm EW_{min}} \simeq 2.5.
\end{eqnarray}

We conclude that the long-term variations in the H$\alpha$ blue-shifted absorption component might suggest a central star undergoing recursive mass ejections in a time interval of about 5000 d. Mass-loss rate appears to change of a factor $\sim 2.5$ $\rm \dot M$. Such a scenario is also supported by the conclusions of \citet{Khalilov2019} for the absorption components of the sodium D lines.

From the above considerations, we can infer on the correlation between long-term mass-loss enhancements and the RV modulation discussed in Section \ref{sec:RVmodulation}. In Fig. \ref{fig:EW} we report, in colored regions, the time intervals where the RV curves are available. The yellow rectangles correspond to high amplitude RV curves (B and E epochs), while gray regions indicate epochs where small amplitude RV curves are found (A, C and D). Small amplitude RV curves correspond to phases of decreasing mass-loss rate, while high amplitude RV curves are in the correspondence of mass-loss enhancements. This piece of evidence might suggest that the stellar regions where the absorption H$\alpha$ component and metal lines are formed, are kinematically linked. 

\input{Tables/Lines_S_PROCESS_Splitting.tex}

\begin{figure*}
\begin{center}
\includegraphics[trim=6.5cm 12cm 0.7cm 2.2cm, clip=true,width=15.2cm,angle=180]{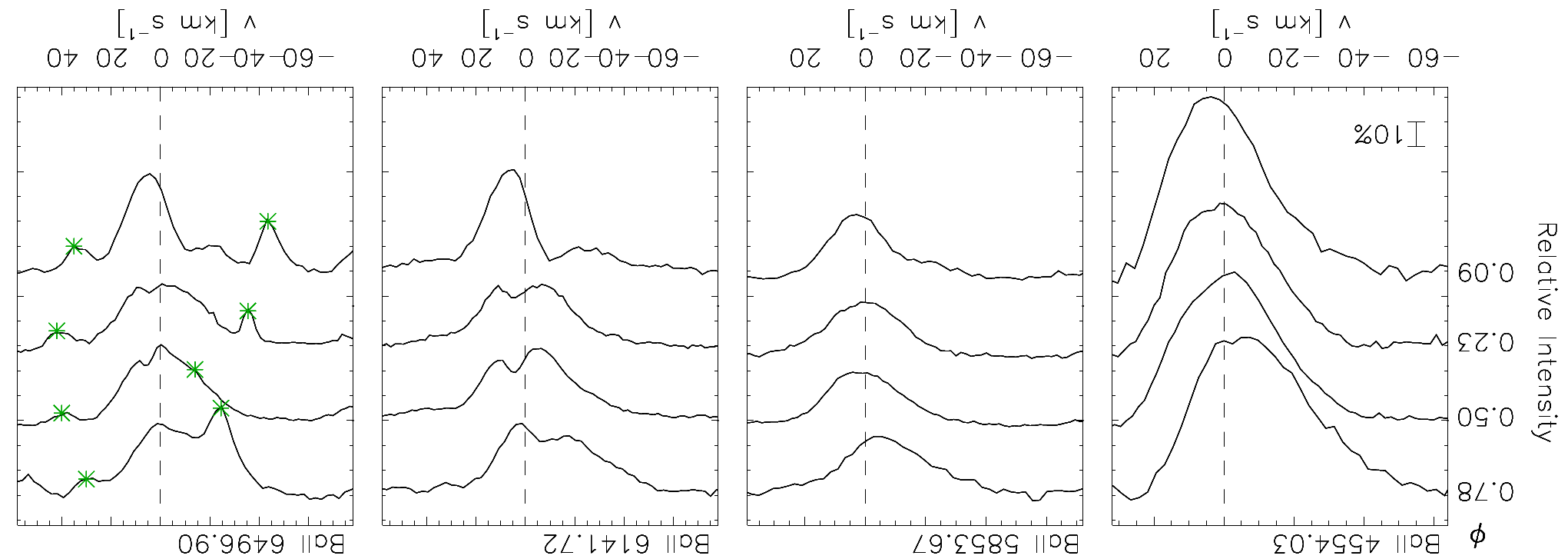}
\includegraphics[trim=6.5cm 12cm 0.7cm 2.2cm, clip=true,width=15.2cm,angle=180]{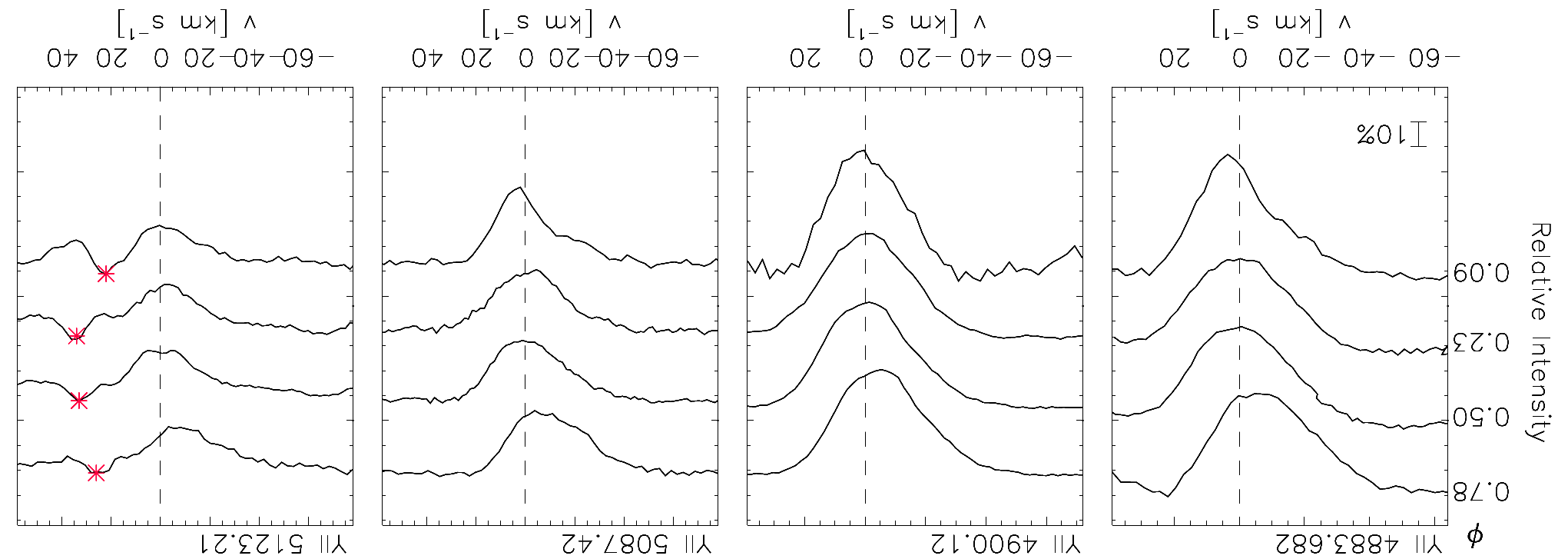}
\begin{center}\caption{\label{fig:S_PROFILES} Observed \ion{Ba}{ii} and \ion{Y}{ii} line profiles for a representative sub-sample of our data, vertically shifted for a better visualization. Profiles are corrected for the RV of the star and ordered according to the orbital phase. Green and red points indicate the contamination from telluric and emission lines, respectively.}\end{center}
\end{center}
\end{figure*}

\begin{figure}
\begin{center}
\includegraphics[trim=5.8cm 8cm 3.2cm 2.2cm, clip=true,width=6cm,angle=180]{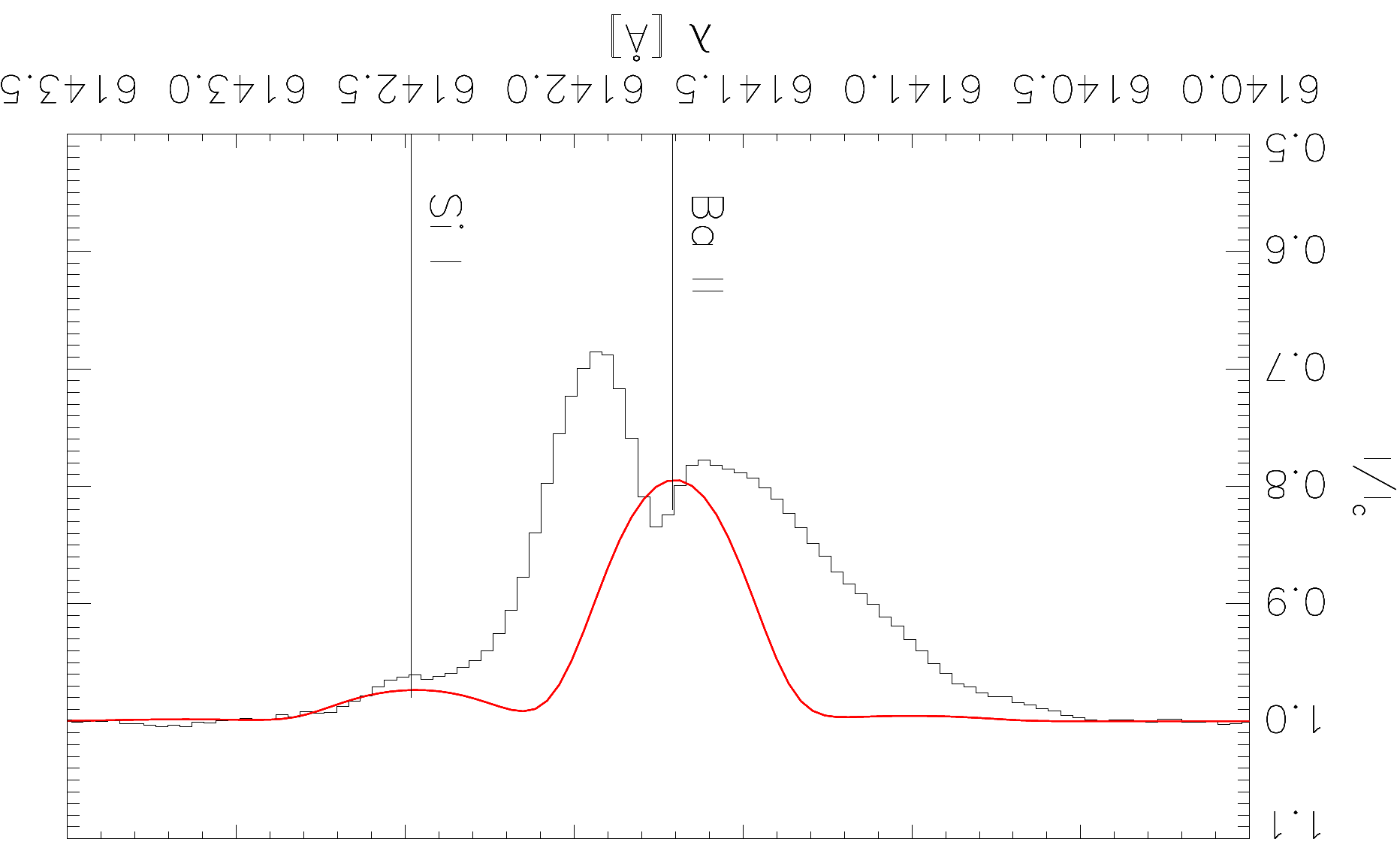}\qquad
\includegraphics[trim=5.8cm 8cm 3.2cm 2.2cm, clip=true,width=6cm,angle=180]{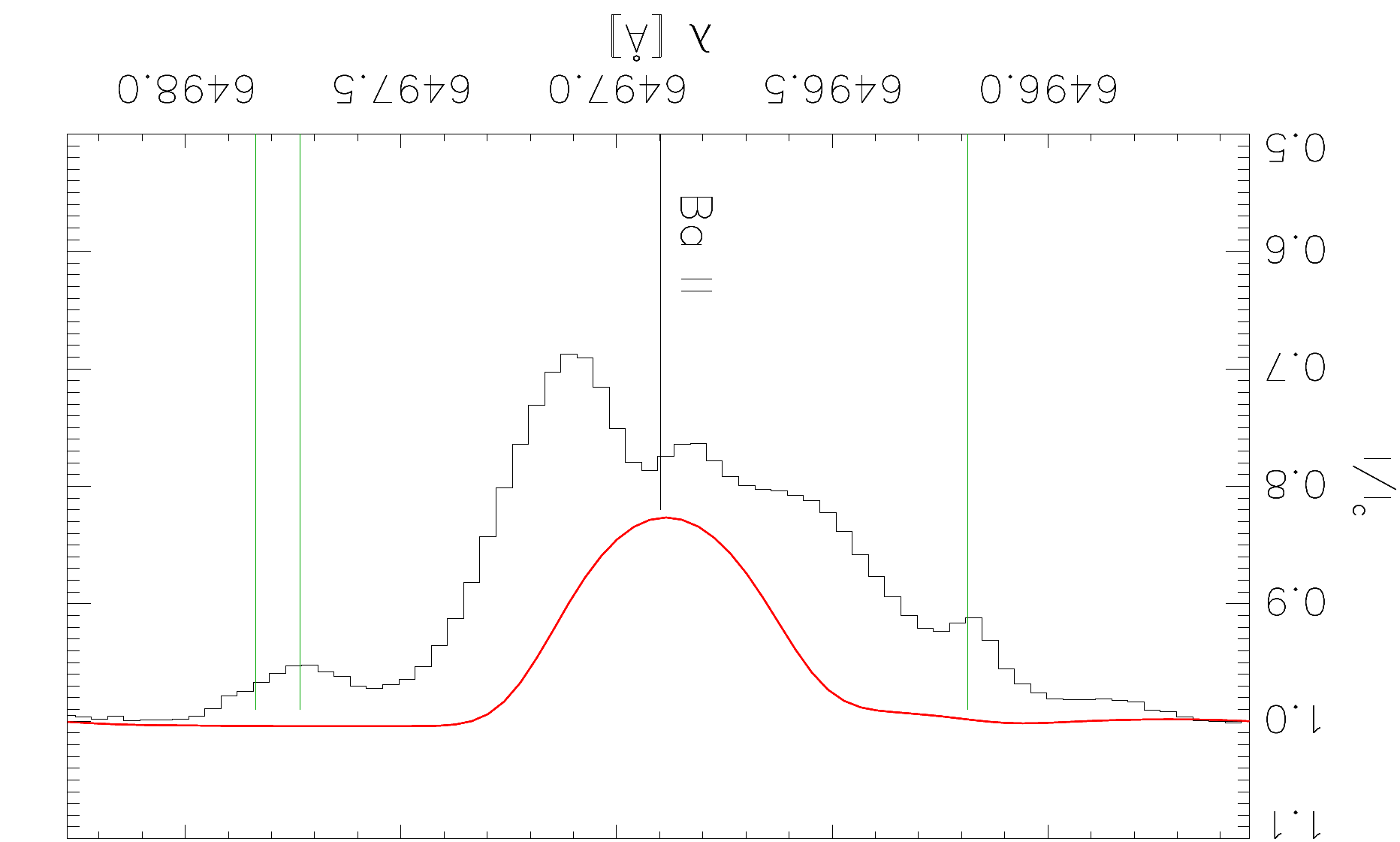}
\begin{center}\caption{\label{fig:BA_SYNTH} \ion{Ba}{ii} $6141.713$ \AA\ and $6496.898$ \AA\ line profiles (black). The solid (red) continuum represents the synthetic profiles computed with \textsc{synthe} \citep{Kurucz1993} adopting the photospheric parameters and abundances by \citet{Kipper2011}. The green lines mark the position of $\rm H_2O$ telluric components. Data were acquired with ESPaDOnS on HJD $53961.777$.}\end{center}
\end{center}
\end{figure}

\section{Short term variability}
\subsection{s-process elements}\label{sec:Selem}
As already mentioned in the Introduction, the interaction between mass-loss and nebula plays a key role for the understanding of the chemical abundance in evolved stars. It is commonly understood the presence of a stable disk necessary to explain the abundance depletion of refractoriness elements, like those produced via slow neutron-capture process. In principle, a spectral analysis of s-process elements could reveal important information about the interaction between central star and the circumstellar environment \citep{Klochkova2014}. Indeed, strong asymmetries or splitting were found in spectral lines of s-process elements. The blue absorption components are associated with structured circumstellar shells, while the red component can be related to photospheric layers. The large variability often observed in such profiles \citep[e.g.][and references therein]{Klochkova2019}, allows us to investigate the instability of the extended atmosphere of evolved stars, characterized by expanding shells of ejected mass and layers falling back on the stellar surface. These processes lead to an effective transport into the envelope of matter synthesized in the previous stage of stellar evolution. Despite its importance, however, a complete characterization of the morphological variability in s-process elements lines is still missing. 

89 Her is found to be under-abundant in s-process elements \citep{Kipper2011} and it has a stable circumbinary disk with an infrared excess compatible with the presence of dusty components \citep{Hillen2013}. These evidences, together with the long-term atmospheric instabilities reported in the previous Sections, make this system a promising candidate for investigating the processes just described. To find out any possible evidence of such processes, we have analyzed the line morphologies in the high resolution spectra of 89 Her. The photospheric metallic lines present slight asymmetries that are compatible with the convective velocities characteristics of the spectral type of the star \citep{Gray1982}. On the contrary, we have found that all lines of s-process elements with low (i.e. $\leq 1$ $\rm eV$) excitation energies of the lower level show strong asymmetry or splitting and, more important, time variability. The list of such elements is reported in Table \ref{tab:SProcLines} while an example of the line morphologies is shown in Fig. \ref{fig:S_PROFILES}. Such figure suggests that the variability in the line shape of the s-process elements could be associated with the orbital period of the system.


 We focus now our analysis on the \ion{Ba}{ii} line profiles at $6141.713$ \AA\ and $6496.898$ \AA, whose components present the largest separation and amplitude of variability. At this purpose,  we performed a Gaussian decomposition of such lines. For each profile we have first identified the photospheric contribution as well as the possible contamination of other photospheric or telluric lines. Fig. \ref{fig:BA_SYNTH} shows a comparison between the observed profiles and the synthetic ones. The latter have been computed with the photospheric parameters and abundances found by \citet{Kipper2011}. In the case of \ion{Ba}{ii} $6141.713$ \AA\ we found that the profile is contaminated by a weak line of \ion{Si}{i} at $6142.487$ \AA, while the \ion{Ba}{ii} $6496.898$ \AA\, line is blended with three telluric lines of $\rm H_2O$ at $6495.862$ \AA, $6497.50$ \AA\ and $6497.594$ \AA. We have then characterized all these contributions before performing the Gaussian decomposition.

\subsubsection{Photospheric contributions}
\begin{figure}
\begin{center}
\includegraphics[trim=1.6cm 0.9cm 8.2cm 17.8cm, clip=true,width=6cm,angle=0]{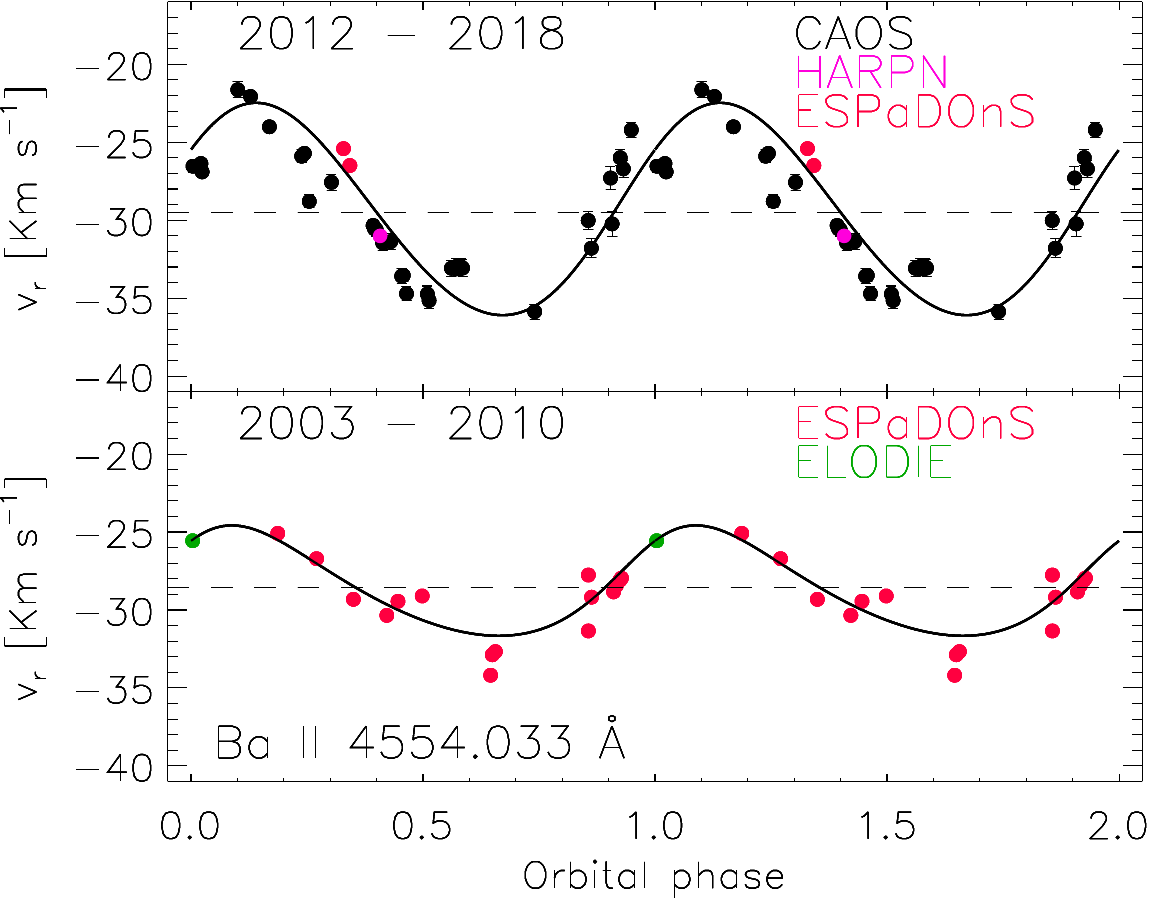}
\includegraphics[trim=1.6cm 0.9cm 8.2cm 17.8cm, clip=true,width=6cm,angle=0]{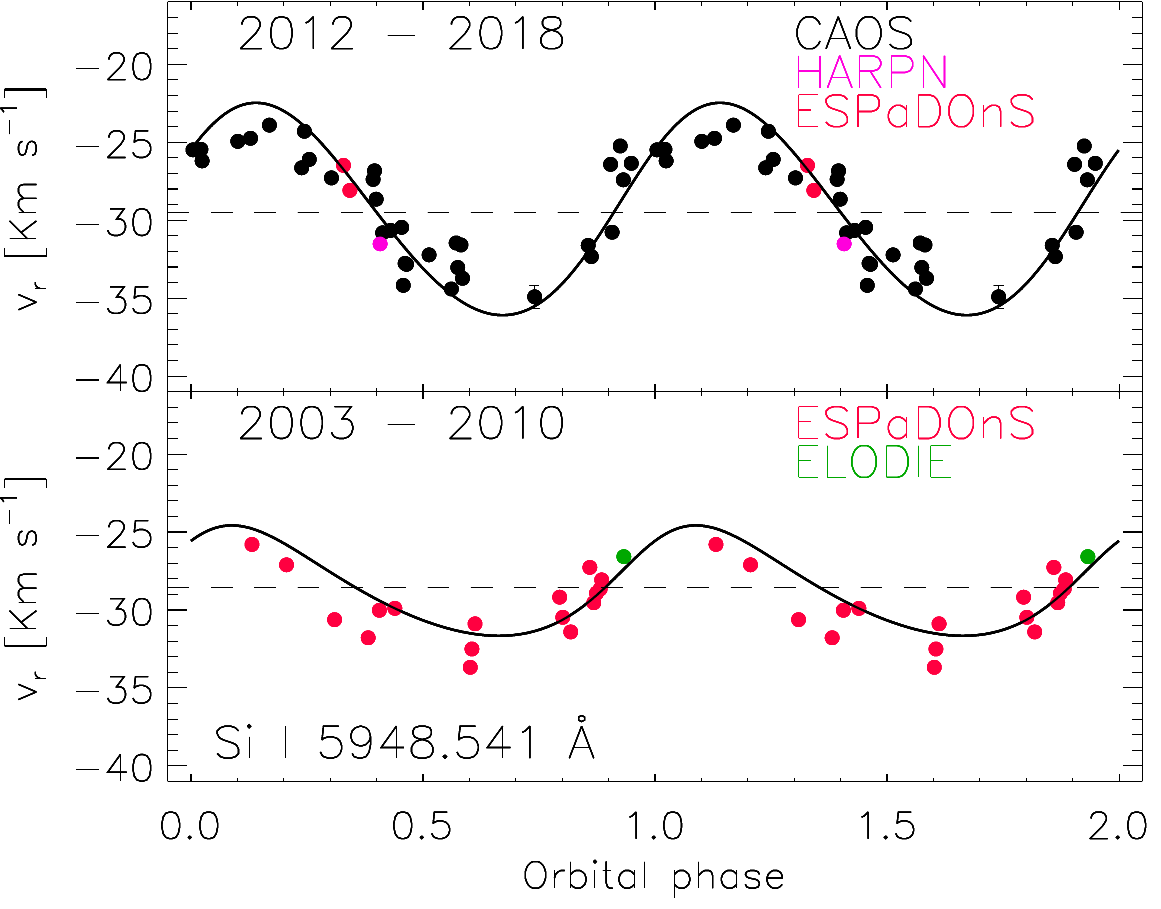}
\begin{center}\caption{\label{fig:RadVelBa} As measured from \ion{Ba}{ii} $4554.033$ \AA\ (top) and the \ion{Si}{i} $5948.541$ \AA\  (bottom) lines, radial velocities are plotted according to orbital phase. Theoretical curves, computed with the orbital parameters reported in Table \ref{tab:OrbitalParameters}, are superimposed.}\end{center}
\end{center}
\end{figure}

The position of the photospheric lines within the morphologies to be fitted can be fixed from the RV calculated in Sec. \ref{sec:RadVel}. In principle in stars with extended and pulsating photospheres, different lines may have different velocities (e.g. \citealt{Kraus2019}). To fit these two \ion{Ba}{ii} lines and the associated \ion{Si}{i} line, we first investigated the unblended and unsplitted lines of \ion{Ba}{ii} $4554.033$ \AA\ and \ion{Si}{i} $5948.541$ \AA. Radial velocities measurements have been carried out following the \emph{center-of-gravity} method \citep{Kochukhov2001}. As shown in Fig. \ref{fig:RadVelBa}, we found that the obtained RV curves are compatible, within errors, with those reported in Sec. \ref{sec:RadVel}. We have then fixed the positions of the photospheric lines following the RV curve of Sec. \ref{sec:RadVel}.

The full-width-at-half-maximum (FWHM) of the photospheric profiles can be fixed by considering: (i) the mean microturbulence velocity $\rm v_t=6.7 \pm 0.3$ $\rm km$ $\rm s^{-1}$ for ionised species and $\rm v_t=3.7 \pm 0.9$ $\rm km$ $\rm s^{-1}$ for neutral atoms \citep{Kipper2011}, (ii) the rotational velocity $\rm v\sin{\rm i} = 23$ $\rm km$ $\rm s^{-1}$ \citep{Hoffleit1991}, (iii) the instrumental broadening. Considering all of these contributions, we have then adopted $\rm FWHM=0.56$ \AA\ and $\rm FWHM=0.49$ \AA\ for the \ion{Ba}{ii} and the \ion{Si}{i} component respectively.

Finally, line strengths can be difficult to estimate. In the case of \ion{Si}{i} $6142.487$ \AA, the abundance reported by \citet{Kipper2011} led to a good fit of the profile (Fig. \ref{fig:BA_SYNTH}). On the contrary for \ion{Ba}{ii} the obtained depth may exceed that of the profile. The photospheric abundance of \ion{Ba}{ii} is then overestimated: this is due to the fact that the equivalent widths (EWs) of \ion{Ba}{ii} have important non-photospheric contributions. We have then fixed the abundance of \ion{Si}{i} to match the observed line strength, while for \ion{Ba}{ii} we set the line strength as a free parameter in the Gaussian Decomposition. 

\subsubsection{Telluric contributions}
As already mentioned, the \ion{Ba}{ii} line profile contains three telluric lines of $\rm H_2O$ at $6495.862$ \AA, $6497.50$ \AA\ and $6497.594$ \AA\, \citep{Moore1966}.

To subtract these contributions, we have fixed the position and the FWHM considering the heliocentric velocity 
and the instrumental broadening respectively. The depth of the lines were computed in the Gaussian decomposition.

\subsubsection{Gaussian decomposition and results}
\begin{figure*}
\begin{center}
\includegraphics[trim=14cm 3.1cm 0.1cm 0cm, clip=true,width=6.7cm,angle=180]{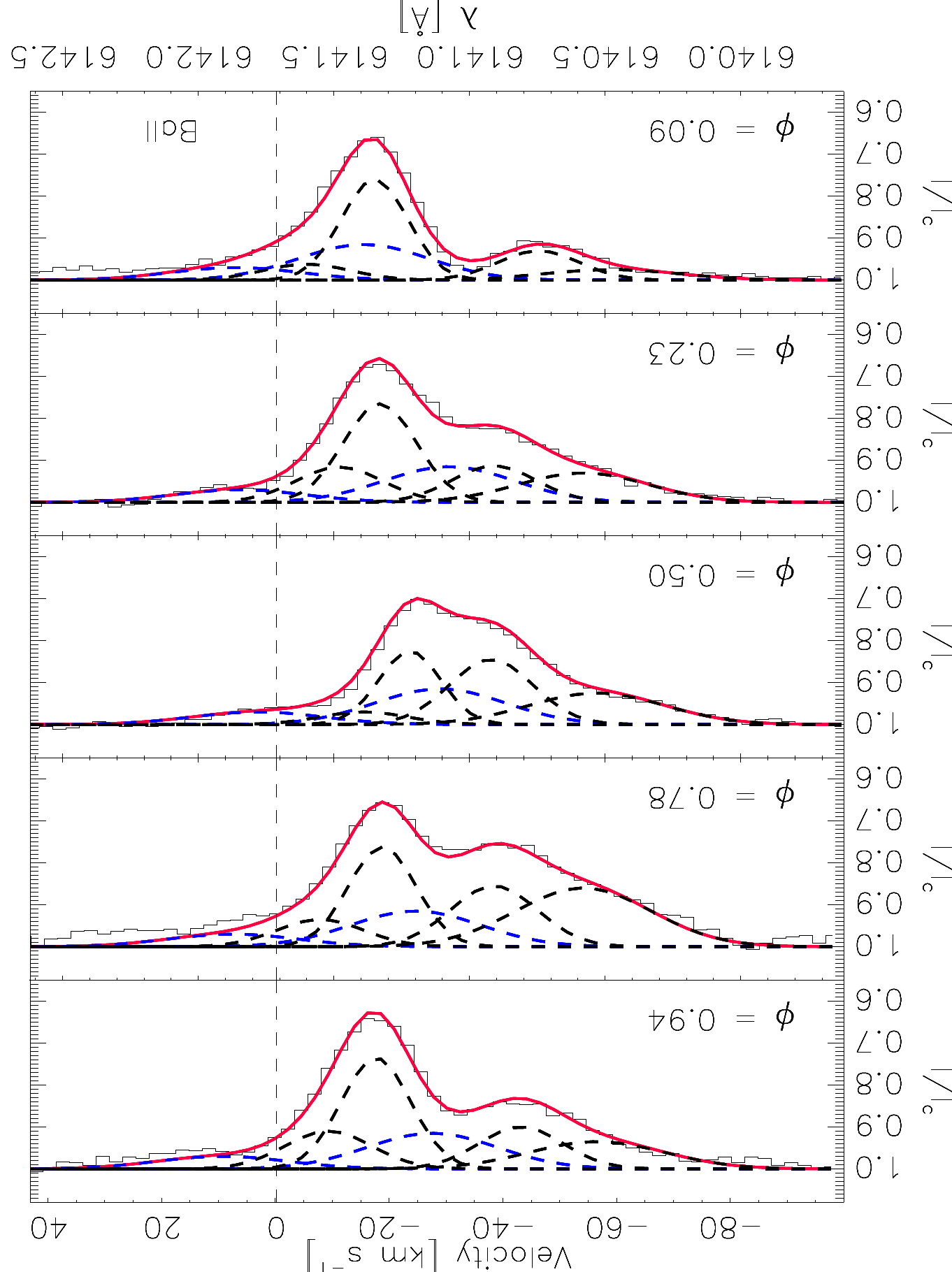}
\includegraphics[trim=14cm 3.1cm 0.1cm 0cm, clip=true,width=6.7cm,angle=180]{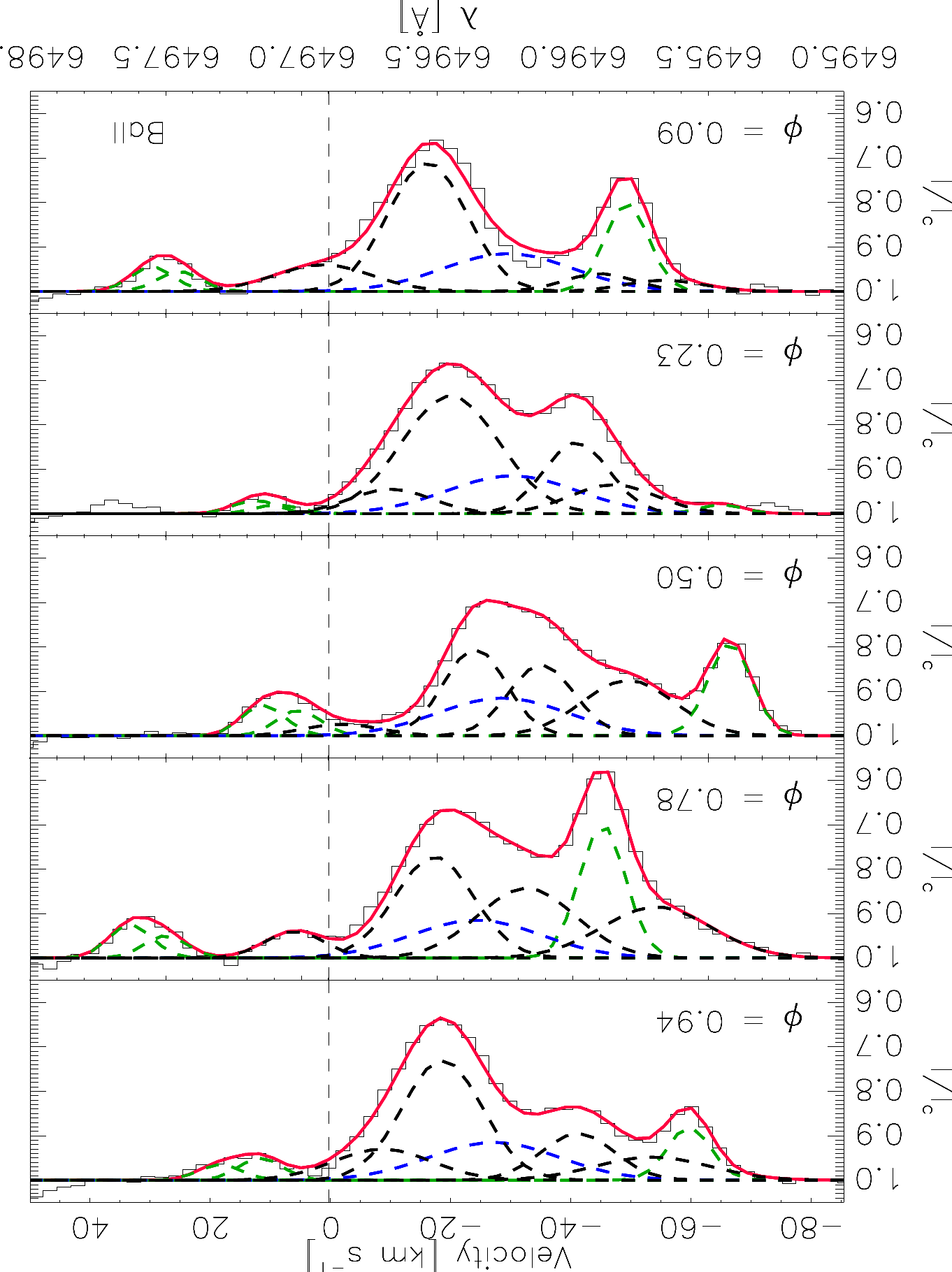}
\begin{center}\caption{\label{fig:BaFit} Example of CAOS-\ion{Ba}{ii} line profiles (continuum black) decomposed with Gaussian components (blue dashed: photospheric, black dashed: circumstellar, green dashed: telluric). The solid (red) continuum is the sum of the components. Each panel is labelled with orbital phase.} \end{center}
\end{center}
\end{figure*}

Gaussian decomposition was performed adopting a line-fitting IDL procedure developed in a previous work to fit multicomponent optical/infrared spectra \citep{Gangi2020}. This procedure is based on $\chi^2$ minimization and provides, for each component, width, peak velocity and peak intensity values. Errors were statistically estimated as the variation in the parameters that increases by 1 the $\chi^2$ value. The total number of components was determined as the minimum number of Gaussians that yields a $\chi^2$ stable at 20\% of its minimum value. In addition to the photospheric and telluric components already mentioned, and for which we have fixed all the parameters except the intensities of the photospheric \ion{Ba}{II} and telluric lines, we found that four other components are necessary to fit the variable \ion{Ba}{ii} $6141.713$ \AA\ and $6496.898$ \AA\ line profiles. Therefore, the fitting procedure determine 3 parameters for each of the four components and the intensities of the photospheric \ion{Ba}{II} and of the telluric contributions. Fig. \ref{fig:BaFit} shows an example of the fit for a representative sub-sample of CAOS data. Despite the periodic variability in radial velocity, two of the four components, hereafter component 0 and 1, were found always red-shifted with respect to the photospheric contribution, while the other two, the components 2 and 3, were found blue-shifted. 

\begin{figure*}
\begin{center}
\includegraphics[trim=9.5cm 9.5cm 0.1cm 0.5cm, clip=true,width=8.9cm,angle=180]{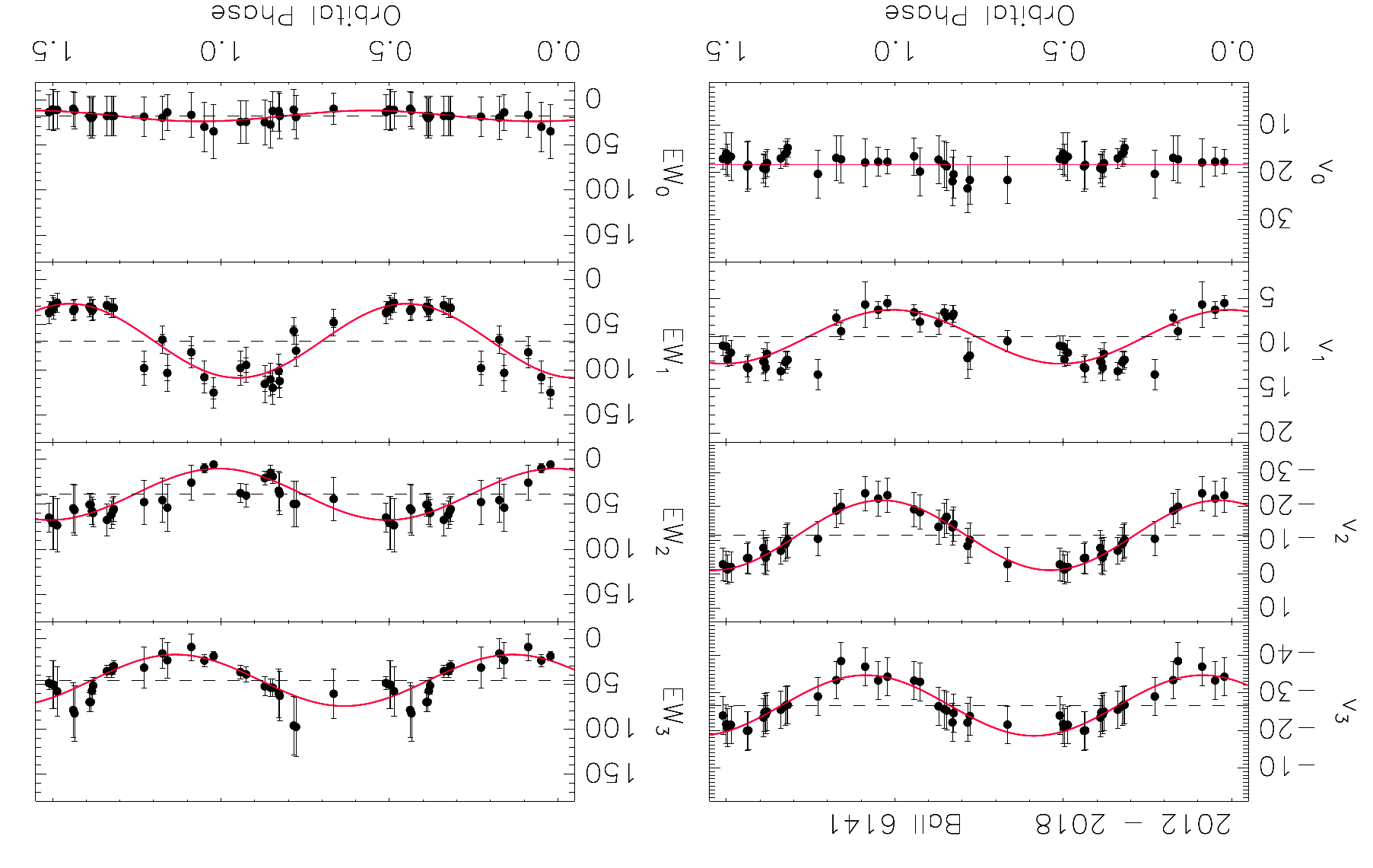}
\includegraphics[trim=9.5cm 9.5cm 0.1cm 0.5cm, clip=true,width=8.9cm,angle=180]{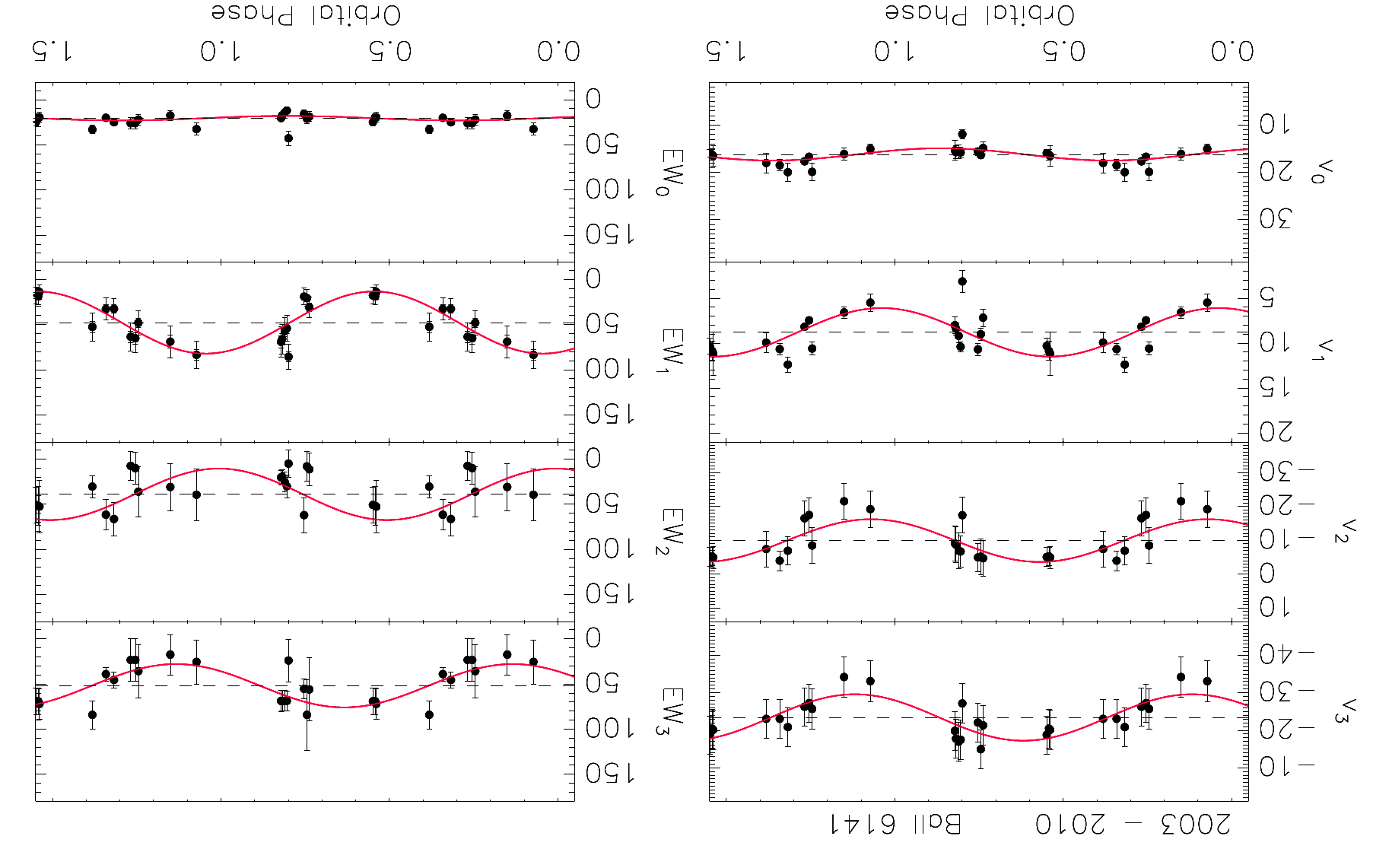}\qquad
\includegraphics[trim=9.5cm 9.5cm 0.1cm 0.5cm, clip=true,width=8.9cm,angle=180]{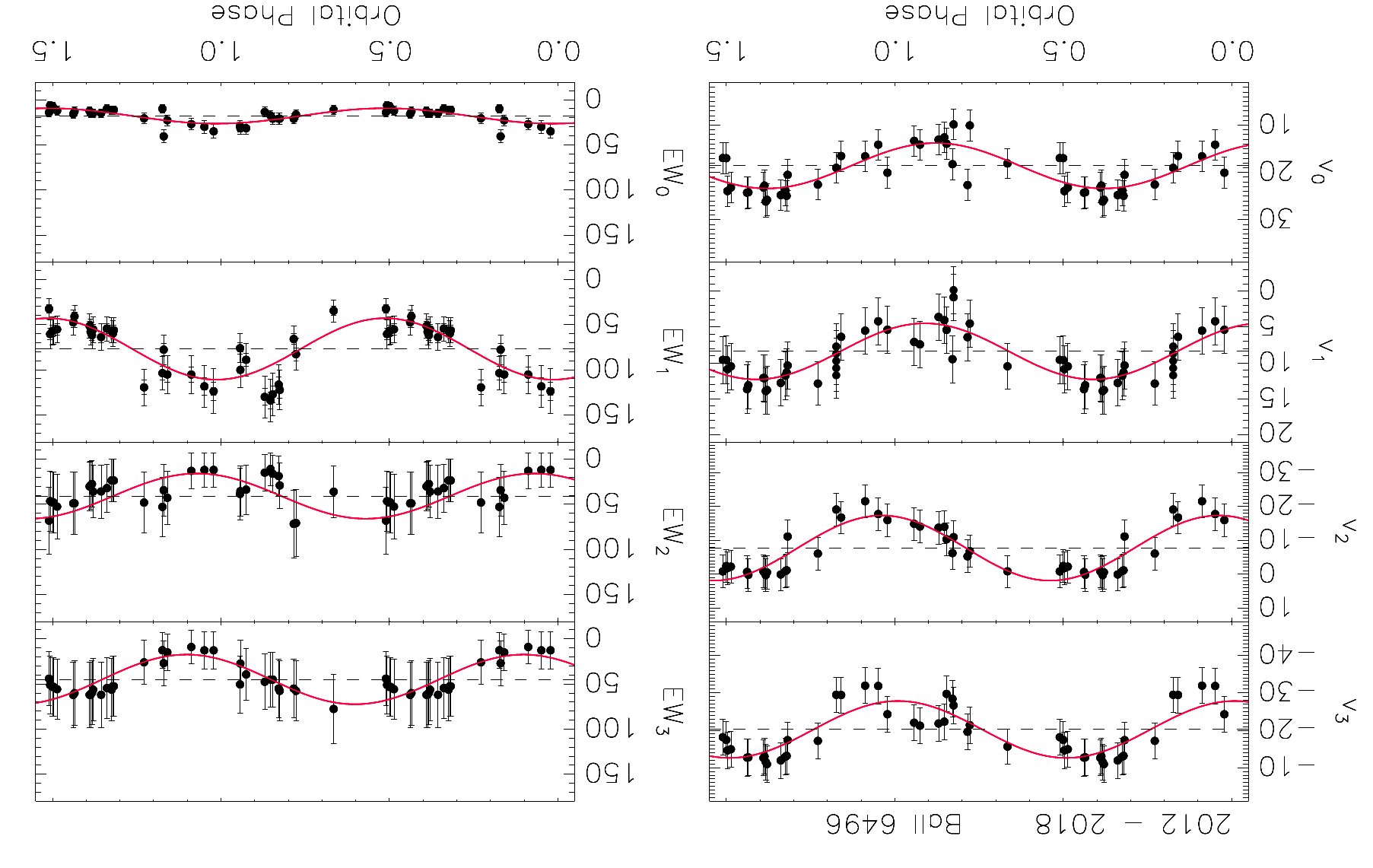}
\includegraphics[trim=9.5cm 9.5cm 0.1cm 0.5cm, clip=true,width=8.9cm,angle=180]{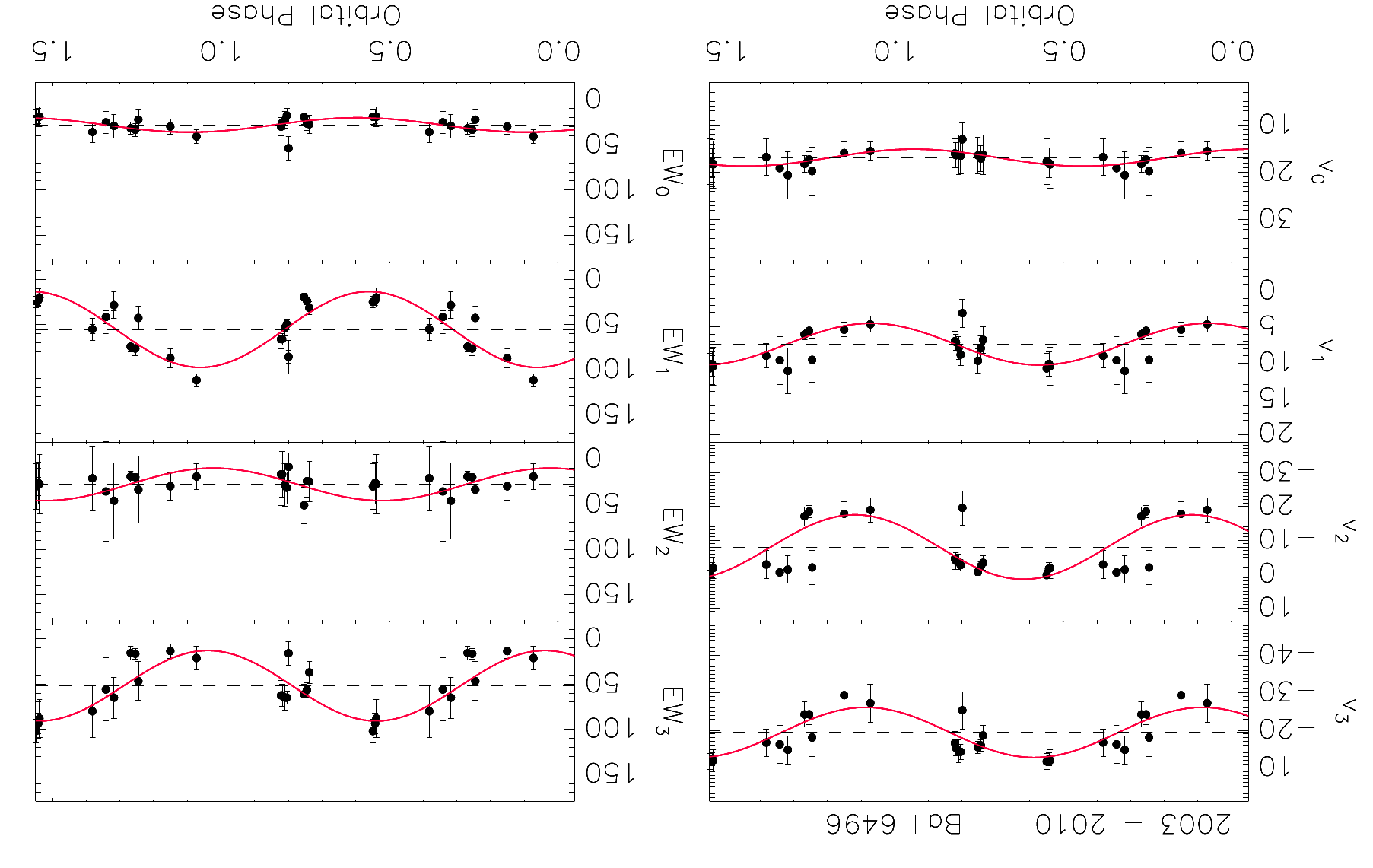}
\begin{center}\caption{\label{fig:ResDec} Gaussian decomposition parameters best matching the \ion{Ba}{ii} 6141.713 \AA\ (top panel) and \ion{Ba}{ii} 6496.898 \AA, lines, folded with the orbital ephemeris given in Table \ref{tab:OrbitalParameters}.} \end{center}
\end{center}
\end{figure*}

Fig. \ref{fig:ResDec} reports the fitted values against the orbital period. Data were divided in the epochs E (2012-2018) and D (2003-2010), to point out any possible different behaviour similar to what was discussed in Section \ref{sec:RadVel}. EWs are given by gaussian-function areas, while peak velocities were subtracted from the photospheric radial velocities. In this ways, the $\rm v_p$ of the four components represent the net velocity with respect to the photosphere of the star. In general, we note that:
\begin{itemize}
\item $\rm v_p$ and EWs show a well-defined variability with the orbital period of the system. The EWs variability is mainly due to strength changes, since we did not find any significant variation in the width of the components.
\item There is no appreciable difference in the behaviour of the fitted parameters between epochs E and D.
\item Compared with other components, the $\rm v_p$ and EWs of component 0 show little significant variability.
\item The velocity curves show always positive (red-shifted) values for the components 0 and 1 and always negative (blue-shifted) values for components 2 and 3.
\item The velocity curves are in phase each other, while they show a $\rm \sim 0.5$ phase difference with the binary radial velocity curve (Fig. \ref{fig:RadVel}). 
\item The EWs curves of component 1 are opposite in phase with respect to those of components 3 and 4.
\end{itemize}

\subsection{Metal emission lines}\label{sec:Emission}
89 Her is known to show numerous permitted low excitation metal lines in emission \citep{Sargent1969, Clim1987, Kipper2011}. Previous works have suggested that the presence of these lines is a proof of collisionally excited interaction between the stellar wind and the circumbinary disk. In addition, some of these lines present linear polarization that can be explained through Thomson scattering in a circumbinary disc of radius $\rm R < 10$ au \citep{Leone2018,Gangi2019}. Such lines could be then considered as a good tracer for investigating possible interaction between the central object and the closer circumstellar regions. 

The emission lines in the optical spectra of 89 Her are quite narrow, showing widths between 8 and 12 $\rm km$ $\rm s^{-1}$ that significantly exceeds only the width of the ESPaDOnS instrumental profile ($\sigma \sim 4$ $\rm km$ $\rm s^{-1}$), if we exclude the single HARPS spectrum. For this reason we limit our analysis to the ESPaDOnS data sub-sample, which are almost all from region D. 

 We have selected transition of metallic lines whose profiles have the highest signal-to-noise ratio (SNR), no blending with other emission or absorption lines and no contamination with telluric lines. We have thus selected 32 lines and, for each orbital phase, we have measured the peak velocity v$\rm _p$, $\rm EW$ and $\rm FWHM$ with a Gaussian fit. The FWHMs were deconvolved by the instrumental width assuming a Gaussian profile. We have found that v$\rm _p$ values are variable with the orbital period of the system, thus indicating that the regions producing the emission are affected by the orbital dynamics, while EWs and FWHMs do not show any appreciable variability. The line list and the average v$\rm _p$, EWs and FWHMs values are reported in Table \ref{tab:emission_lines}.

\begin{figure}
\begin{center}
\includegraphics[trim= 2.cm 0.6cm 7.6cm 9cm, clip=true,angle=0,width=5.6cm]{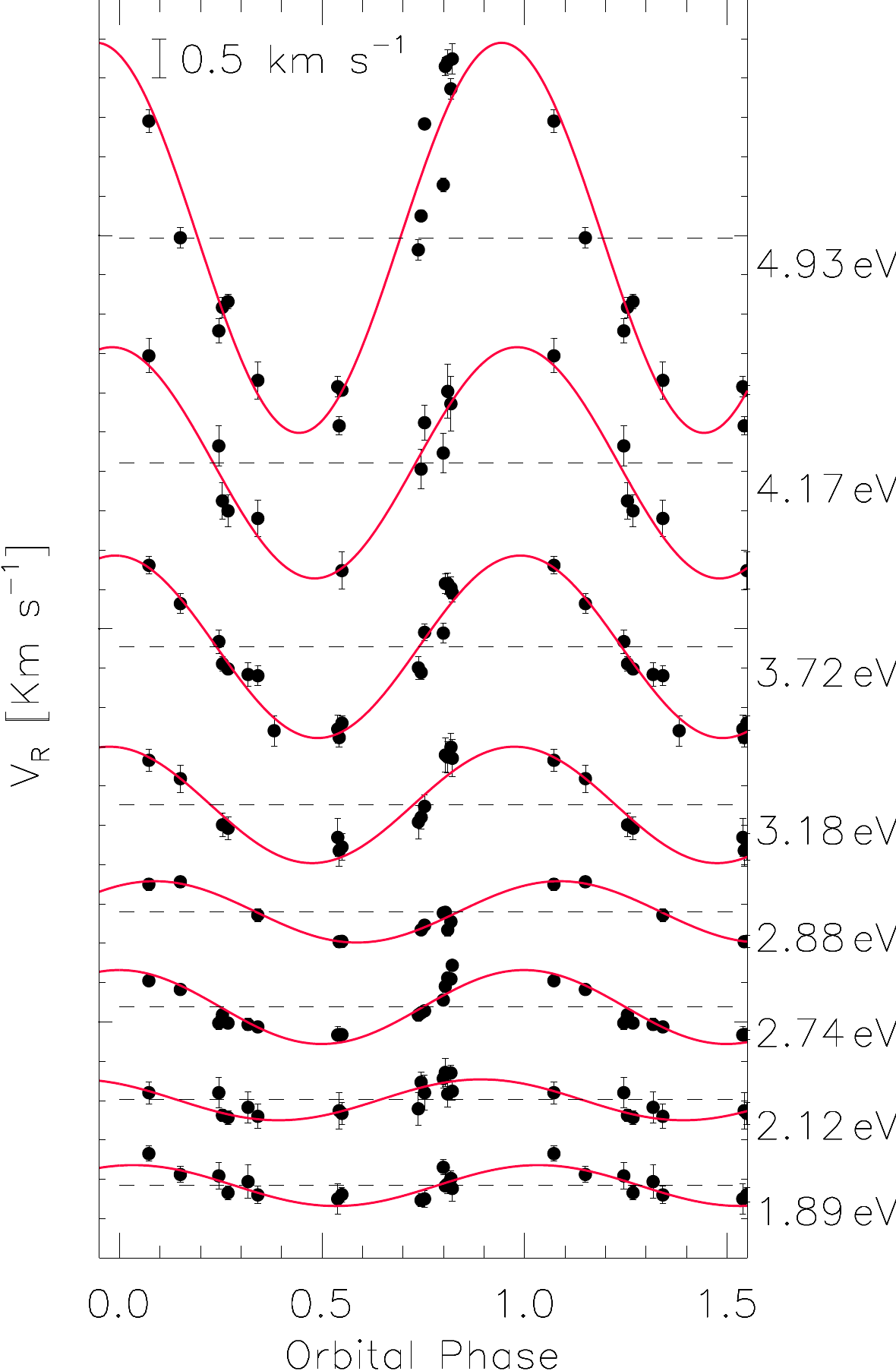}
\caption{\label{fig:VelShift} Radial velocity measurements for a representative sub-sample of emission lines, plotted according to the orbital phase and vertically shifted for a better visualization. The amplitude of the variations increases with increasing energy of the upper level transition (labelled in the figure for each curve). Sinusoidal fit (in red) are also plotted.}
\end{center}
\end{figure}

\begin{figure}
\begin{center}
\includegraphics[trim= 2.4cm 8.cm 4.2cm 0cm, clip=true,angle=180,width=8cm]{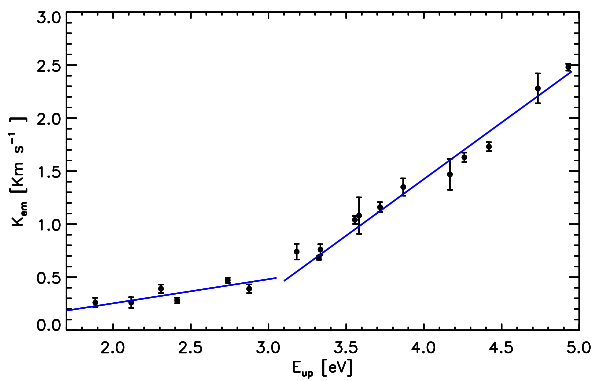}
\caption{\label{fig:Results_fit_seno} Semi-amplitude of the velocity curves $\rm K_{em}$ of emission lines plotted against the upper level transition energies $\rm E_{up}$. Linear fits are reported in blue lines.}
\end{center}
\end{figure}

\begin{figure}
\begin{center}
\includegraphics[trim= 8cm 8cm 2.5cm 0.5cm, clip=true,angle=180,width=7.5cm]{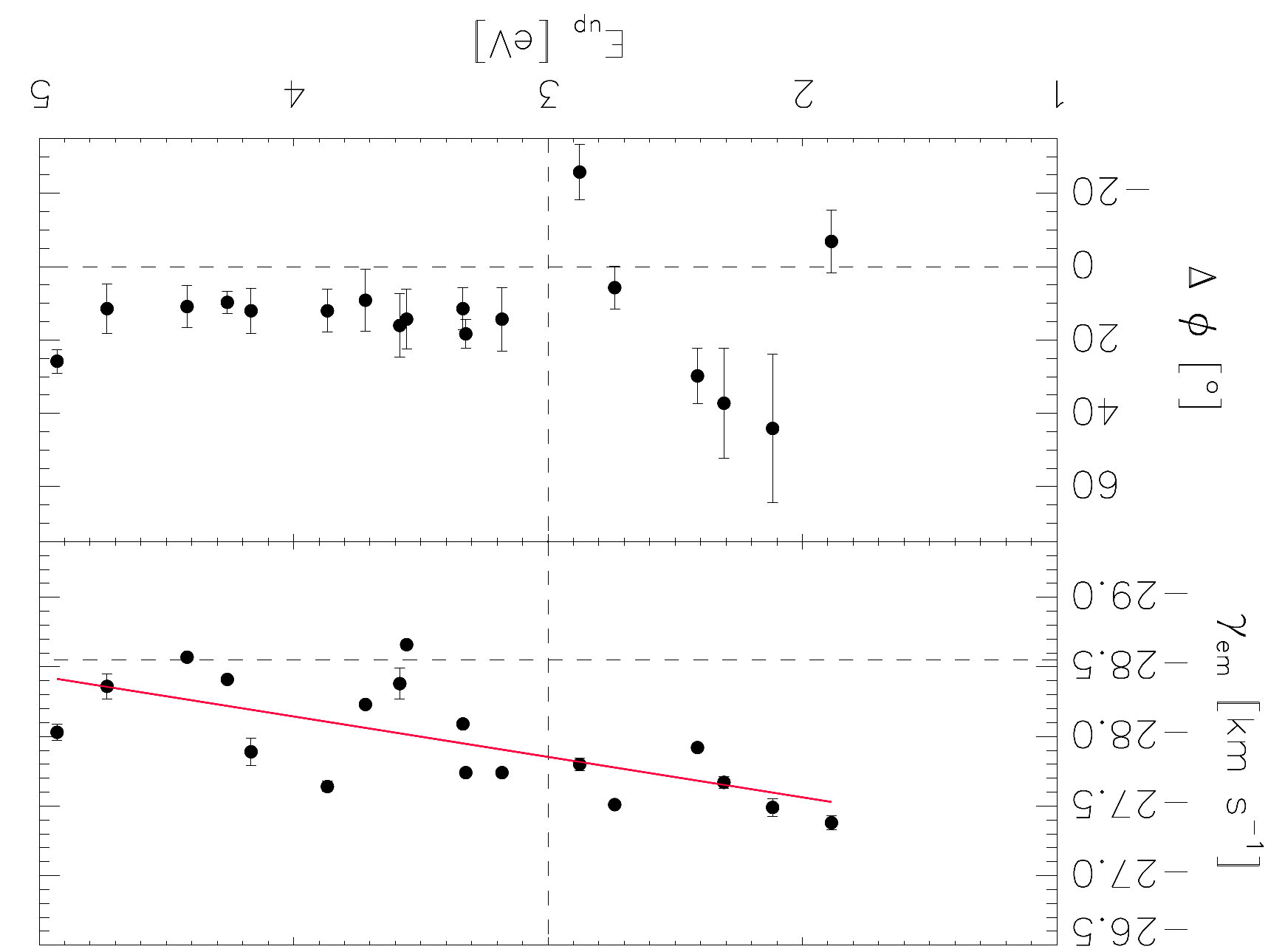}
\caption{\label{fig:Results_fit_seno_vel} Top: average velocity $\rm \gamma_{em}$ plotted against the upper level transition energies $\rm E_{up}$. Horizontal dashed line represents the value of the systemic velocity $\gamma$ of the system. Bottom: RV curves phase difference between emission and photospheric lines ($\rm \Delta \phi = \phi_{em} - \phi_{abs}$).}
\end{center}
\end{figure}

Looking at the radial velocities, we have found that the amplitude of the curves $\rm k_{em}$ is smaller than that measured for the photospheric absorption lines and strongly dependent on the energy of the upper level transitions $\rm E_{up}$ (Fig. \ref{fig:VelShift}). We then measured the semi-amplitude of the RV curves $\rm k_{em}$ with a sinusoidal fit of the data, averaging RV values corresponding to lines whose upper energies differ for less than $\rm 0.02\, eV$, to further increase the SNR. Fig. \ref{fig:Results_fit_seno} shows the strong correlation found between $\rm E_{up}$ and $\rm k_{em}$. The semi-amplitude of the RV curves scales linearly with $\rm E_{up}$ and a change in the slope at energies $\rm E_{up} < 3 eV$ is also observed. Fig. \ref{fig:Results_fit_seno_vel} (top) reports the average velocity $\rm \gamma_{em}$ calculated from the sinusoidal fit against $\rm E_{up}$. A slight linear increase of about 1 $\rm km$ $\rm s^{-1}$ is observed going towards higher $\rm E_{up}$. Values corresponding at $\rm E_{up} > 3 eV$ are compatible, within errors, with the systemic velocity ($\rm \gamma=-28.5 \pm 0.5$). Fig. \ref{fig:Results_fit_seno_vel} (bottom) shows the RV curves phase difference $\rm \Delta \phi$ between emission and photospheric lines. A constant $\rm \Delta \phi = 10^{\circ}$ is observed in lines with $\rm E_{up} > 3\, eV$ while a large variation up to $\rm \Delta \phi \sim 40^{\circ}$ is found for $\rm E_{up} < 3\, eV$.

\section{Discussion}\label{sec:Discussion}
In the following we discuss the obtained results in terms of long-term (i.e. in time scale $\rm \tau$ greater than the orbital period $\rm P$) and short-term (i.e. with $\rm \tau \simeq P$ ) variations.

\subsection{Long-term variations}
We have found long-term variations in the optical spectra of 89 Her, in terms of RV curves and H$\alpha$ line morphology. In particular, a correlation between the RV amplitude and the H$\alpha$ $\rm EW_{abs}$ might suggest a common origin.

To better support this scenario we also looked at possible long-term variations in the brightness of the source. Fig. \ref{fig:Photometry} shows the time series of available magnitude $\rm V_{mag}$ of 89 Her as measured by the American Association of Variable Star Observers (AAVSO). As already reported in previously works, short-term variations up to $\sim$ 0.2 mag are ascribable to the pulsational and binary nature of the system. Looking at the bottom panel of Fig. \ref{fig:Photometry}, in the years 2004/2011 (epoch D) there is a slightly constant increase in the $\rm V_{mag}$ up to $\sim 0.06$ mag, while in the years 2012/2020 (epoch E) a constant decrease is observed. Therefore, it would appear that, at least for periods D and E, there is a correlation between H$\alpha$ $\rm EW_{abs}$, RV amplitude and stellar brightness.

In the following we discuss two possible scenarios.

\subsubsection{Orbital wobbling}
As discussed in Sec. \ref{sec:RVmodulation}, the observed RV amplitude modulation could be compatible with an orbital wobbling. Focusing in particular to the epochs D and E, we found that a variation of the orbital disk inclination of about $10^\circ$ would lead to the observed increase in the RV amplitude.
In this scenario, the behaviour of the H$\alpha$ $\rm EW_{abs}$ could be understood in terms of misalignment between the axis of the expanding hour-glass structure, supposed to be perpendicular to the orbital plane, and the line-of-sight direction. When the hour-glass structure is aligned with the observer direction, the terminal velocity $\rm v_{\infty}$ of the H$\alpha$ absorption component reaches the maximum value, while as the misalignment increases the line-of-sight intercepts only the upper edges of the jet, with a slower expansion velocity.

We have found that epoch D is characterized by an important decrease in the $\rm v_{\infty}$ of the H$\alpha$ absorption component from about $-250$ $\rm km$ $\rm s^{-1}$ to about $-100$ $\rm km$ $\rm s^{-1}$, while epoch E starts with a minimum value of $\rm v_{\infty}$ of about $-70$ $\rm km$ $\rm s^{-1}$ that increases up to about $-190$ $\rm km$ $\rm s^{-1}$ (Fig. \ref{fig:EW}). In this context, we could suppose that during epoch D, the orbital plane inclination, with a medium value of about $10^\circ$, increases up to a certain maximum value corresponding to the beginning of epoch E. The latter would be then characterized by decreasing orbital plane inclination with a medium value of about $20^\circ$.

However this scenario would be not compatible with the trend of the visual photometry reported in Fig. \ref{fig:Photometry}. Indeed, we would expect an increase in the brightness of the source during epoch D, because it would be less obscured from the optically thick expanding shell, and an increase during epoch E. Instead, an opposite behaviour is observed.

\subsubsection{Recursive mass-ejection events}
Alternatively, the observed variations in the blue-shifted absorption component of the P Cygni-like H$\alpha$ profile might suggest the existence of recursive mass-ejection events occurring every $\sim 5000$\,d (i.e. $\sim 13.7$ years). Considering the geometry of the circumstellar environment, with a circumbinary dusty disk and an expanding hour-glass structure, it is highly probable that during the an "eruptive" event the material is ejected along the direction of the hour-glass axis. In the subsequent "relaxation" phase, the optically thick ejected shell would be dispersed in the circumstellar environment, becoming less dense and no more visible as a blue-shifted absorption component. Indeed, the H$\alpha$ line never shows an inverse P Cygni profile, indicating that an infall phase of the ejected material does not occur.
 
 In such a scenario, the correlation between H$\alpha$ $\rm EW_{abs}$ and the RV amplitudes could be explained in terms of dynamics within the extended atmosphere of the star and the RV amplitude variations would no longer be related with the binary orbital parameters. Indeed, during the "eruptive" event, atmospheric velocity gradient perturbations could cause an expansion of the stellar atmosphere that, persisting for a time interval greater than the orbital period, would leads to an increase in the amplitude of the RV curve. On the contrary the "relaxation" phase would be characterized by a compression of the atmosphere with a decrease in the RV amplitude curve. In addition, the fact that the asymmetry of the photospheric absorption lines does not shows any appreciable variability between the eruptive and relaxation phases (see Fig. \ref{fig:BIS_4508}), might also suggest that such expansion is not limited to the upper layers of the photosphere.
 
\begin{figure*}
\begin{center}
\includegraphics[trim= 1cm 19.5cm 6cm 0.5cm, clip=true,width=15cm]{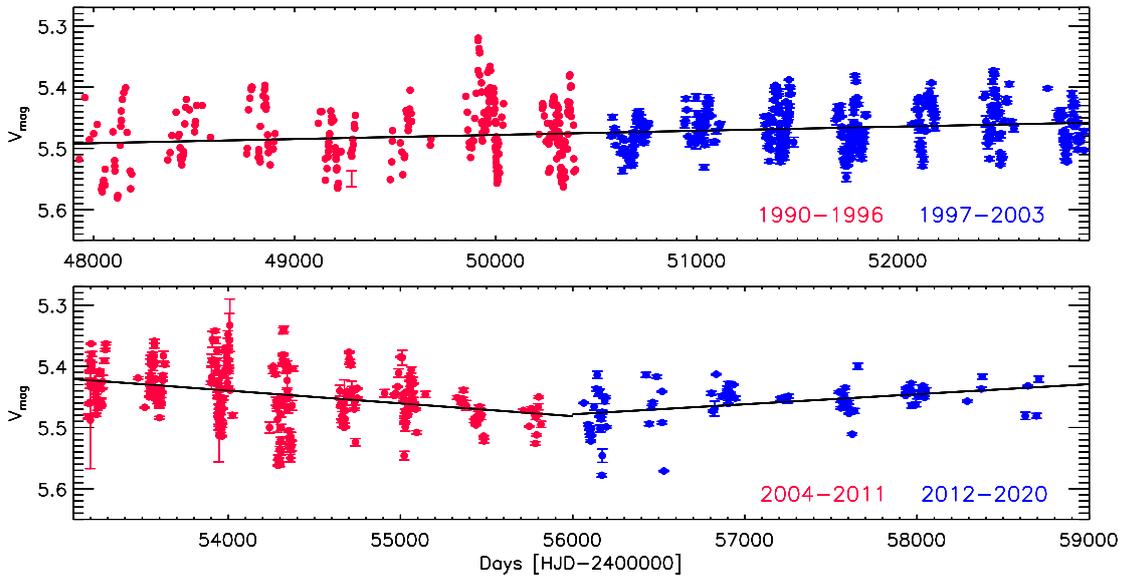}
\caption{\label{fig:Photometry} Visual magnitude of 89 Her as measured by AAVSO during the years 1990-2003 (top) and 2004-2020 (bottom). The minimum of the luminosity around JD=2456000 is coincident with the $\rm H_{\alpha}$ minimum.}
\end{center}
\end{figure*}

 Finally, a variation of the emitting surface due to the photospheric expansion or compression, would lead to a variation of the stellar brightness in line with the behaviour reported in Fig. \ref{fig:Photometry}.

Atmospheric eruptive episodes of mass loss are known among evolved and massive yellow hypergiants with observed time scales of one to a few decades \citep[e.g.][]{Genderen2019}. Such episodes are characterized by very strong mass-loss rates that can reach values of the order of $\rm 10^{-2}$ $\rm M_{\sun}$ $\rm yr^{-1}$, as predicted  by hydrodynamical models \citep{Tuchman1978,Stothers1999}. An example is provided by the yellow hypergiant star $\rm \rho$ Cas, where the mass-loss rate estimated during the mass-ejection episode in the years 2000/2001 amounted to $\sim 3 \cdot 10^{-2}$ $\rm M_{\sun}$ $\rm yr^{-1}$ \citep{Lobel2003} and a new episode of mass-ejection was registered $\sim 12$ years later \citep{Kraus2019}, a result in line with the predictions. Similar results have also been predicted for more evolved and lower mass (i.e. $\rm M < M_{\sun}$) stars, like the AGBs \citep[e.g.][]{Wood1974}. 

The time interval between the two supposed episodes of mass loss in 89 Her would appears to be compatible with that observed and predicted for massive yellow hypergiants and AGBs. Despite the important differences in the amount of mass loss, this scenario might suggests a possible similarity in the underlying physics for these kinds of processes, providing a link between the late evolutionary phases of stars and PNe formation.

\subsection{Short-term variations}
We have found variations in the spectra of 89 Herculis that cover time scales compatible with the orbital period of the system, for which we refer to as short-term variations.

The first of these concerns the morphological variability in the lines corresponding to s-process elements, in terms of strong blue-wing asymmetries or splitting. Focusing to the \ion{Ba}{ii} at $6141.713$ \AA\ and $6496.898$ \AA, a gaussian-function decomposition allowed us to identify, for each profile, four components having $\rm v_p$ and EWs variable with the orbital period of the system. These components could identify shells of processed material that is moving with respect to the star. In this scenario, the behaviour of the peak velocity shown in Fig. \ref{fig:ResDec} suggests that the absorption components 2 and 3 may represent ejected shells that are moving toward us, and that their dynamics is affected by the orbital motion of the system. On the contrary, components 0 and 1 are representative of in-falling material. The lower variability seen in the component 0 may also suggest that the location of the corresponding shell is far away from the central star and it is then not very influenced by the orbital dynamics. This idea is also supported by looking at the behaviour of $\rm v_0$ of the $6496.898$ \AA\, line profile during the epoch D (2004-2010) and E (2014-2018): the slight increase in amplitude in the epoch E suggests that the position of the in-falling shell is now close to the system and therefore is more affected by orbital motion. In addition, the phase differences between the binary RV curve and the \ion{Ba}{ii} RV components of about 0.5 P, suggest that such circumstellar components are subjected to the orbital motion of the secondary star. Finally, the behaviour of the EWs may be explained in terms of geometrical effects, if inhomogeneities within the shells exist.

These kind of splitting have already been observed in evolved stars, although their temporal variability has never been completely investigated. In the already mentioned case of the hypergiant $\rm \rho$ Cas, the variability observed in the splitting of the strong \ion{Ba}{ii} 6141 \AA\, line was associated with the one decade cadence kinematic instabilities of the upper atmospheric layers \citep{Klochkova2019}. In our case we do not see any correlation between splitting variability of s-process elements and the long-term changes of the star. 

The second kind of short-term variations we found is related to the kinematic of the metal lines in emission, which show RV curves variable with the orbital period of the system. The most relevant observed evidence is the dependence of the amplitude of the RV curves with the energy of the upper level transition. This may suggest that the regions producing the emission are not stationary with respect to the photosphere of the primary component and that lines are formed in regions with different sizes. Indeed, lines corresponding to higher $\rm E_{up}$ could form in regions close to the star, thus becoming more influenced by the binary motion, while those with lower $\rm E_{up}$ could form in larger regions and thus be no more strongly subjected to the binary orbit.

The behaviour of $\rm \gamma_{em}$ and $\rm \Delta \phi$ observed in Fig. \ref{fig:Results_fit_seno_vel} may be an indication of the presence of a complex structure and dynamics within the close circumstellar disk. Indeed, the slight increase toward low $\rm E_{up}$ of the $\rm \gamma_{em}$ velocity may suggest the decrease in the expanding velocity from the inner to the outer regions. In particular, by subtracting the systemic velocity of the system ($\rm \gamma = -28.5 \pm 0.5$), it emerges that external regions of the disk may fall in toward the internal regions. The phase shift between the emission and the absorption photospheric lines is of about $\rm 12^\circ$ for lines with $\rm E_{up} > $ $\rm 3 eV$, while it increase up to $\sim 40^\circ$ for lines with $\rm E_{up} < $ $\rm 3 eV$. This behaviour may be explained if the structure of the disk present some type of large scale asymmetries which lead to regions of enhancement emission located at different angles respect to the direction of the semi-major axis of the binary orbit. Such a scenario, however, needs further high-quality data and suitable models to be better investigated.

\section{Summary and conclusion}\label{sec:Conclusion}
In this paper we have presented a long-term optical spectral monitoring of the post-AGB binary system 89 Herculis. The main results of our study can be summarized as follows:
\begin{itemize}
    \item Radial velocity curve of the system shows long-term changes, mostly in amplitude. Such variations cannot be related to long-term cycle-to-cycle pulsation variations, presence of a third body in a hierarchical triple star system and mass transfer between the two components of the system, but they can be compatible with an orbital plane wobbling.
    \item The P Cygni like H$\alpha$ profile shows long-term changes with important increases in the EWs of the blue-shifted absorption component in a time interval of about 5000 d. This variability is correlated with that seen in the amplitudes of the RV curves, with high RV amplitude corresponding to increasing EWs.
    \item Two different scenarios are proposed to explain the above pieces of evidence: (i) the RV amplitude modulation is due to orbital plane wobbling and the H$\alpha$ $\rm EW_{abs}$ variations are interpreted in terms of misalignment between the axis of the hour-glass structure, perpendicular to the orbital plane, and the line-of-sight; (ii) the RV amplitude modulation is due to velocity photospheric perturbations that are associated with recursive mass-ejection episodes which explain the H$\alpha$ $\rm EW_{abs}$ variations.
    \item Absorption lines corresponding to s-process elements show important asymmetries or splitting. The splitting observed in the \ion{Ba}{ii} $6141.713$ \AA\ and $6496.898$ \AA\ line profiles could be explained by four shell components, two expanding and two infalling, subject to the orbital motion of the secondary star. No long-term effects are observed.
    \item Metal lines in emission show radial velocity variations according to the orbital period of the system with amplitudes proportional to the upper energy level $\rm E_{up}$ of the transition. A complex dynamics, in term of falling-in material and large scale disk asymmetries may also be present. 
\end{itemize}

Our results highlight the importance of high resolution spectroscopy over a wide temporal range to study the interaction between photospheric instabilities and dynamics in the close envelopes of evolved stars. At the same time, we note the need for suitable theoretical models able to correctly interpret the observational evidences.

\section*{Acknowledgements}
This research used also the facilities of the Canadian Astronomy Data Centre operated by the National Research Council of Canada with the support of the Canadian Space Agency. Partially based on observations made with the Italian Telescopio Nazionale Galileo (TNG) operated on the island of La Palma by the Fundacion Galileo Galilei of the INAF (Istituto Nazionale di Astrofisica) at the Spanish Observatorio del Roque de los Muchachos of the Instituto de Astrofisica de Canarias. We acknowledge the support by INAF/Frontiera through the "Progetti Premiali" funding scheme of the Italian Ministry of Education, University, and Research. We also acknowledge financial contribution from the agreement ASI-INAF n.2018-16-HH.0 and from \emph{Programma ricerca di Ateneo UNICT 2020-22 linea 2}. M. Gangi expresses gratitude to Dr. S. Antoniucci, K. Biazzo, T. Giannini and B. Nisini for helpful comments and a critical read of the paper. Finally, we thank the anonymous referee for the significant constructive comments.

\section*{Data Availability}
The data underlying this article will be shared on reasonable request to the corresponding author.


\bibliographystyle{mnras}
\bibliography{Bibliography}


\appendix

\section{Additional plots and tables}

\input{Tables/table_emission_lines.tex}


\bsp	
\label{lastpage}
\end{document}

%% file: Tables/RadialVelocities.tex
\begin{table*}
\small
\caption{\label{tab:RadVel} Journal of observations and radial velocity measurements. OBS: \textbf{RIT} Ritter, \textbf{ELO} Elodie, \textbf{ESP} Espadons, \textbf{LRS} DOLORES, \textbf{FER} Feros, \textbf{CAOS} Caos, \textbf{HAN} Hanpo, \textbf{HAR} HARPS-N, \textbf{BTA} Nasmyth Echelle Spectrograph of the BTA telescope.}
\begin{tabular}{ccclcccl}
 \hline
 \hline
Date      & HJD        & RV            & OBS & Date      & HJD        & RV            & OBS  \\
day-mo-yr  & [-2400000] & $\rm km$ $\rm s^{-1}$ &     & day-mo-yr  & [-2400000] & $\rm km$ $\rm s^{-1}$ &      \\
 \hline
  24-09-1993 & 49254.520 & -26.86 $\pm$ 0.22 & RIT    &      04-07-2014 & 56843.397 & -31.09 $\pm$ 0.18 & CAOS \\  
  01-04-1994 & 49443.883 & -29.27 $\pm$ 0.18 & RIT    &      05-07-2014 & 56844.437 & -31.32 $\pm$ 0.23 & CAOS \\     
  11-05-1994 & 49483.800 & -24.26 $\pm$ 0.16 & RIT    &      09-07-2014 & 56848.446 & -32.32 $\pm$ 0.15 & CAOS \\ 
  22-05-1994 & 49494.792 & -22.88 $\pm$ 0.22 & RIT    &      14-07-2014 & 56853.442 & -31.81 $\pm$ 0.17 & CAOS \\       
  29-05-1994 & 49501.745 & -26.16 $\pm$ 0.27 & RIT    &      21-07-2014 & 56860.320 & -33.31 $\pm$ 0.28 & CAOS \\      
  07-06-1994 & 49510.773 & -24.67 $\pm$ 0.16 & RIT    &      22-07-2014 & 56861.319 & -34.61 $\pm$ 0.18 & CAOS \\       
  01-07-1995 & 49899.800 & -27.43 $\pm$ 0.20 & RIT    &      23-07-2014 & 56862.463 & -34.00 $\pm$ 0.18 & CAOS \\       
  31-07-1995 & 49929.679 & -33.02 $\pm$ 0.23 & RIT    &      24-07-2014 & 56863.354 & -34.41 $\pm$ 0.17 & CAOS \\     
  04-07-1996 & 50268.691 & -28.54 $\pm$ 0.28 & RIT    &      25-07-2014 & 56864.317 & -34.10 $\pm$ 0.23 & CAOS \\      
  23-05-1997 & 50591.774 & -29.29 $\pm$ 0.31 & RIT    &      06-08-2014 & 56876.348 & -35.42 $\pm$ 0.20 & CAOS \\       
  20-07-1997 & 50649.664 & -25.69 $\pm$ 0.19 & RIT    &      07-08-2014 & 56877.436 & -35.50 $\pm$ 0.30 & CAOS \\ 
  05-04-1998 & 50908.896 & -26.56 $\pm$ 0.41 & RIT    &      21-08-2014 & 56891.344 & -34.41 $\pm$ 0.17 & CAOS \\
  12-04-2003 & 52741.570 & -29.60 $\pm$ 0.25 & BTA    &      24-08-2014 & 56894.317 & -35.17 $\pm$ 0.17 & CAOS \\
  18-08-2004 & 53236.300 & -24.83 $\pm$ 0.43 & ELO    &      25-08-2014 & 56895.317 & -34.21 $\pm$ 0.24 & CAOS \\
  22-08-2005 & 53604.753 & -25.05 $\pm$ 0.22 & ESP    &      27-08-2014 & 56897.308 & -34.79 $\pm$ 0.31 & CAOS \\
  08-02-2006 & 53775.115 & -28.17 $\pm$ 0.21 & ESP    &      28-08-2014 & 56898.283 & -35.35 $\pm$ 0.20 & CAOS \\
  10-02-2006 & 53777.096 & -29.67 $\pm$ 0.17 & ESP    &      03-09-2014 & 56904.321 & -34.60 $\pm$ 0.22 & CAOS \\   
  14-08-2006 & 53961.777 & -29.38 $\pm$ 0.17 & ESP    &      10-10-2014 & 56941.000 & -34.98 $\pm$ 0.19 & CAOS \\ 
  02-03-2007 & 54162.066 & -23.47 $\pm$ 0.26 & ESP    &      12-10-2014 & 56943.000 & -35.32 $\pm$ 0.12 & CAOS \\ 
  29-09-2007 & 54372.821 & -23.30 $\pm$ 0.21 & ESP    &      16-06-2016 & 57555.522 & -29.60 $\pm$ 0.19 & CAOS \\ 
  14-02-2009 & 54877.099 & -31.83 $\pm$ 0.17 & ESP    &      17-06-2016 & 57557.489 & -31.36 $\pm$ 0.20 & CAOS \\ 
  15-02-2009 & 54878.132 & -31.51 $\pm$ 0.17 & ESP    &      29-06-2016 & 57569.489 & -25.93 $\pm$ 0.19 & CAOS \\ 
  17-02-2009 & 54880.076 & -31.22 $\pm$ 0.11 & ESP    &      30-06-2016 & 57570.442 & -27.40 $\pm$ 0.20 & CAOS \\ 
  02-05-2009 & 54954.125 & -28.63 $\pm$ 0.19 & ESP    &      06-07-2016 & 57575.513 & -24.25 $\pm$ 0.16 & CAOS \\ 
  04-05-2009 & 54955.849 & -28.39 $\pm$ 0.19 & ESP    &      07-07-2016 & 57577.455 & -23.82 $\pm$ 0.18 & CAOS \\ 
  06-05-2009 & 54958.121 & -27.59 $\pm$ 0.17 & ESP    &      12-07-2016 & 57582.411 & -24.30 $\pm$ 0.16 & CAOS \\ 
  07-05-2009 & 54959.076 & -27.20 $\pm$ 0.14 & ESP    &      28-07-2016 & 57598.420 & -25.35 $\pm$ 0.17 & CAOS \\ 
  07-09-2009 & 55081.874 & -29.80 $\pm$ 0.22 & ESP    &      02-08-2016 & 57603.385 & -24.55 $\pm$ 0.17 & CAOS \\ 
  28-09-2009 & 55102.832 & -31.34 $\pm$ 0.16 & ESP    &      25-08-2016 & 57626.387 & -22.29 $\pm$ 0.18 & CAOS \\ 
  05-10-2009 & 55109.823 & -29.86 $\pm$ 0.19 & ESP    &      02-09-2016 & 57634.335 & -23.28 $\pm$ 0.17 & CAOS \\
  01-02-2010 & 55229.169 & -30.87 $\pm$ 0.14 & ESP    &      04-10-2016 & 57666.248 & -25.68 $\pm$ 0.17 & CAOS \\
  05-04-2010 & 55292.470 & -24.47 $\pm$ 0.29 & BTA    &      20-05-2017 & 57893.559 & -25.54 $\pm$ 0.26 & CAOS \\ 
  02-06-2012 & 56080.575 & -34.25 $\pm$ 0.35 & LRS    &      26-05-2017 & 57900.102 & -24.83 $\pm$ 0.30 & HAN  \\ 
  09-05-2013 & 56422.404 & -26.63 $\pm$ 0.49 & FER    &      01-07-2017 & 57935.528 & -23.43 $\pm$ 0.16 & CAOS \\ 
  15-06-2014 & 56824.035 & -27.15 $\pm$ 0.10 & ESP    &      25-07-2017 & 57960.374 & -25.88 $\pm$ 0.15 & CAOS \\ 
  19-06-2014 & 56828.007 & -28.11 $\pm$ 0.11 & ESP    &      13-05-2018 & 58251.592 & -25.16 $\pm$ 0.22 & CAOS \\ 
  22-05-2014 & 56799.619 & -23.43 $\pm$ 0.23 & CAOS   &      29-05-2018 & 58267.548 & -27.29 $\pm$ 0.20 & CAOS \\
  08-06-2014 & 56816.519 & -31.71 $\pm$ 0.17 & CAOS   &      25-06-2018 & 58294.013 & -33.31 $\pm$ 0.18 & HAR  \\
  04-07-2014 & 56842.519 & -31.05 $\pm$ 0.19 & CAOS   &      01-08-2018 & 58332.357 & -36.25 $\pm$ 0.21 & CAOS \\   
 \hline
 \hline
\end{tabular}
\end{table*}

%% file: Tables/OrbitalParameters_NEW_VERSION.tex
\begin{table*}
\small
\center
\caption{\label{tab:OrbitalParameters} Orbital parameters derived through modeling of the radial velocity curves shown in Fig. \ref{fig:RadVel}.}
\begin{tabular}{llccccc}
\hline
\hline
 Epochs                      & & \textbf{A}         & \textbf{B}         & \textbf{C}         & \textbf{D}         & \textbf{E}         \\
                                  & & 1978-1979 & 1982-1989 & 1990-1998 & 2003-2010 & 2012-2018 \\
 \hline
$\rm P$       & $\rm [d]$          & $288.92$ $\pm$ $3.2$   & $288.46$ $\pm$ $1.70$	& $290.42$ $\pm$ $1.90$	& $290.81$ $\pm$ $0.20$	& $289.43$ $\pm$ $0.1$ \\
$\rm T$       &                    & $46150$ $\pm$ $15$   & $46054$ $\pm$ $6$	& $46030$ $\pm$ $17$	& $45965$ $\pm$ $7$	& $46044$ $\pm$ $4$ \\
$\rm e$       &                    & $0.17$ $\pm$ $0.27$   & $0.14$ $\pm$ $0.34$	& $0.17$ $\pm$ $0.22$	& $0.17$ $\pm$ $0.19$	& $0.06$ $\pm$ $0.12$ \\
$\rm \gamma$  & [$\rm km$ $\rm s^{-1}$] & $-27.91$ $\pm$ $0.62$   & $-28.95$ $\pm$ $0.95$	& $-28.74$ $\pm$ $0.46$	& $-28.55$ $\pm$ $0.52$	& $-29.51$ $\pm$ $0.58$ \\
$\rm \omega$  & $\rm [deg]$   & $493$ $\pm$ $34$   & $405$ $\pm$ $14$	& $406$ $\pm$ $22$	& $316$ $\pm$ $12$	& $304$ $\pm$ $7$ \\
$\rm K$       & [$\rm km$ $\rm s^{-1}$] & $2.24$ $\pm$ $0.74$   & $4.76$ $\pm$ $0.66$ & $2.96$ $\pm$ $0.68$ & $3.54$ $\pm$ $0.5$ & $6.81$ $\pm$ $0.77$ \\
\hline
$\rm a sin i$ & [au]               & $0.06$ $\pm$ $0.04$ & $0.12$ $\pm$ $0.06$ & $0.08$ $\pm$ $0.04$ & $0.09$ $\pm$ $0.03$ & $0.18$ $\pm$ $0.04$ \\
$\rm f$       & [$\rm 10^{-4}$ $\rm M_{\sun}$]  & $3.37$ $\pm$ $3.37$ & $32.25$ $\pm$ $13.60$ & $7.81$ $\pm$ $5.43$ & $13.37$ $\pm$ $5.67$ & $94.74$ $\pm$ $32.17$ \\
\hline
\hline
\end{tabular}
\end{table*}

%% file: Tables/Logbook_EW_ABS_EM.tex
\begin{table*}
\small
\caption{\label{tab:ew} Equivalent widths of the absorption ($\rm EW_{ab}$) and emission ($\rm EW_{em}$) components of the $\rm H\alpha$ spectral line. The terminal velocity $\rm v_{\infty}$ of the blue-shifted absorption component is also reported. OBS: \textbf{RIT} Ritter, \textbf{ELO} Elodie, \textbf{ESP} Espadons, \textbf{LRS} DOLORES, \textbf{FER} Feros, \textbf{CAOS} Caos, \textbf{HAN} Hanpo, \textbf{HAR} HARPS-N, \textbf{BTA} Nasmyth Echelle Spectrograph of the BTA telescope.}  
\begin{tabular}{ccccc|ccccc}
\hline
\hline
 HJD       & OBS     & $\rm EW_{ab}$  & $\rm EW_{em}$    & $\rm v_{\infty}$ & HJD      & OBS & $\rm EW_{ab}$ & $\rm EW_{em}$ & $\rm v_{\infty}$ \\
 2400000+  &         & [m\AA]         & [m\AA]           & [$\rm km$ $\rm s^{-1}$] & 2400000+ &     & [m\AA]        & [m\AA]        & [$\rm km$ $\rm s^{-1}$]  \\
\hline
49254.520  &  RIT    & 2101 $\pm$ 215 & 615    $\pm$ 19  &   -149  & 56843.397  &  CAOS   & 1631 $\pm$ 258 & 782    $\pm$ 34    & -98   \\
49443.882  &  RIT    & 1654 $\pm$ 147 & 785    $\pm$ 13  &   -121  & 56844.437  &  CAOS   & 1613 $\pm$ 223 & 755    $\pm$ 31    & -98   \\
49483.800  &  RIT    & 1991 $\pm$ 184 & 699    $\pm$ 73  &   -159  & 56848.446  &  CAOS   & 1633 $\pm$ 220 & 746    $\pm$ 39    & -97   \\   
49494.800  &  RIT    & 2186 $\pm$ 199 & 609    $\pm$ 109 &   -174  & 56853.442  &  CAOS   & 1673 $\pm$ 229 & 751    $\pm$ 69    & -98   \\  
49501.741  &  RIT    & 2143 $\pm$ 170 & 526    $\pm$ 30  &   -174  & 56860.320  &  CAOS   & 1731 $\pm$ 182 & 787    $\pm$ 48    & -95   \\   
49510.800  &  RIT    & 2327 $\pm$ 229 & 601    $\pm$ 91  &   -182  & 56861.319  &  CAOS   & 1712 $\pm$ 295 & 813    $\pm$ 52    & -98   \\   
49899.800  &  RIT    & 2373 $\pm$ 227 & 375    $\pm$ 158 &   -183  & 56862.463  &  CAOS   & 1772 $\pm$ 346 & 623    $\pm$ 42    & -102  \\  
49929.676  &  RIT    & 2248 $\pm$ 385 & 426    $\pm$ 167 &   -176  & 56863.354  &  CAOS   & 1676 $\pm$ 254 & 833    $\pm$ 60    & -95   \\  
50268.691  &  RIT    & 1592 $\pm$ 135 & 849    $\pm$ 89  &   -120  & 56864.317  &  CAOS   & 1798 $\pm$ 255 & 762    $\pm$ 43    & -101  \\    
50591.774  &  RIT    & 1321 $\pm$ 154 & 927    $\pm$ 66  &   -110  & 56876.348  &  CAOS   & 1543 $\pm$ 234 & 823    $\pm$ 24    & -88   \\    
50649.664  &  RIT    & 1691 $\pm$ 269 & 809    $\pm$ 172 &   -121  & 56877.436  &  CAOS   & 1619 $\pm$ 235 & 783    $\pm$ 36    & -96   \\ 
50908.896  &  RIT    & 1810 $\pm$ 365 & 813    $\pm$ 103 &   -135  & 56891.344  &  CAOS   & 1501 $\pm$ 271 & 886    $\pm$ 20    & -85   \\
52741.570  &  BTA    & 2244 $\pm$ 93  & 283    $\pm$ 84  &   -172  & 56894.317  &  CAOS   & 1480 $\pm$ 234 & 892    $\pm$ 48    & -86   \\   
53236.300  &  ELO    & 1660 $\pm$ 153 & 612    $\pm$ 46  &   -122  & 56895.317  &  CAOS   & 1463 $\pm$ 219 & 876    $\pm$ 21    & -89   \\   
53604.753  &  ESP    & 3182 $\pm$ 284 & 949    $\pm$ 17  &   -246  & 56897.308  &  CAOS   & 1525 $\pm$ 214 & 873    $\pm$ 31    & -99   \\   
53775.115  &  ESP    & 3111 $\pm$ 194 & 991    $\pm$ 34  &   -227  & 56898.283  &  CAOS   & 1449 $\pm$ 236 & 903    $\pm$ 46    & -99   \\   
53777.096  &  ESP    & 3100 $\pm$ 233 & 944    $\pm$ 37  &   -228  & 56904.321  &  CAOS   & 1548 $\pm$ 182 & 931    $\pm$ 26    & -108  \\   
53961.777  &  ESP    & 3182 $\pm$ 370 & 1148   $\pm$ 97  &   -228  & 56941.000  &  CAOS   & 1620 $\pm$ 234 & 777    $\pm$ 89    & -106  \\   
54162.066  &  ESP    & 2737 $\pm$ 290 & 1053   $\pm$ 60  &   -214  & 56943.000  &  CAOS   & 1690 $\pm$ 335 & 853    $\pm$ 115   & -107  \\  
54372.821  &  ESP    & 2406 $\pm$ 255 & 959    $\pm$ 36  &   -199  & 57555.522  &  CAOS   & 1985 $\pm$ 351 & 1055   $\pm$ 161   & -128  \\ 
54877.099  &  ESP    & 1968 $\pm$ 285 & 475    $\pm$ 101 &   -145  & 57557.489  &  CAOS   & 1723 $\pm$ 412 & 1082   $\pm$ 142   & -124  \\ 
54878.132  &  ESP    & 1919 $\pm$ 335 & 480    $\pm$ 87  &   -144  & 57569.489  &  CAOS   & 1957 $\pm$ 296 & 1192   $\pm$ 60    & -130  \\ 
54880.076  &  ESP    & 1933 $\pm$ 280 & 500    $\pm$ 119 &   -144  & 57570.442  &  CAOS   & 1853 $\pm$ 402 & 1176   $\pm$ 91    & -127  \\ 
54954.125  &  ESP    & 1489 $\pm$ 223 & 655    $\pm$ 123 &   -113  & 57575.513  &  CAOS   & 1862 $\pm$ 292 & 1151   $\pm$ 125   & -129  \\  
54955.849  &  ESP    & 1475 $\pm$ 284 & 630    $\pm$ 90  &   -112  & 57577.455  &  CAOS   & 1902 $\pm$ 244 & 1021   $\pm$ 87    & -130  \\ 
54958.121  &  ESP    & 1478 $\pm$ 265 & 634    $\pm$ 105 &   -112  & 57582.411  &  CAOS   & 2034 $\pm$ 358 & 1018   $\pm$ 113   & -133  \\ 
54959.076  &  ESP    & 1482 $\pm$ 265 & 647    $\pm$ 116 &   -113  & 57598.420  &  CAOS   & 1738 $\pm$ 326 & 1042   $\pm$ 128   & -125  \\  
55081.874  &  ESP    & 2370 $\pm$ 295 & 589    $\pm$ 75  &   -174  & 57603.385  &  CAOS   & 1743 $\pm$ 316 & 1055   $\pm$ 124   & -126  \\  
55102.832  &  ESP    & 2188 $\pm$ 308 & 729    $\pm$ 59  &   -157  & 57626.387  &  CAOS   & 2229 $\pm$ 366 & 943    $\pm$ 157   & -147  \\  
55109.823  &  ESP    & 2198 $\pm$ 286 & 666    $\pm$ 50  &   -152  & 57634.335  &  CAOS   & 2275 $\pm$ 319 & 907    $\pm$ 188   & -153  \\  
55229.169  &  ESP    & 1557 $\pm$ 237 & 644    $\pm$ 57  &   -111  & 57666.248  &  CAOS   & 2725 $\pm$ 424 &        ...         & -160  \\  
55292.470  &  BTA    & 1497 $\pm$ 92  & 619    $\pm$ 32  &   -118  & 57893.559  &  CAOS   & 2280 $\pm$ 247 & 743    $\pm$ 20    & -149  \\  
56080.575  &  LRS    & 1283 $\pm$ 109 & 475    $\pm$ 29  &   -112  & 57900.102  &  HAN    & 2185 $\pm$ 496 & 759    $\pm$ 161   & -157  \\  
56422.404  &  FER    & 1345 $\pm$ 214 & 377    $\pm$ 52  &   -83   & 57935.528  &  CAOS   & 2233 $\pm$ 378 & 871    $\pm$ 135   & -152  \\  
56824.035  &  ESP    & 1481 $\pm$ 254 & 440    $\pm$ 44  &   -98   & 57960.374  &  CAOS   & 2480 $\pm$ 325 & 784    $\pm$ 34    & -163  \\  
56828.007  &  ESP    & 1467 $\pm$ 268 & 447    $\pm$ 55  &   -98   & 58251.592  &  CAOS   & 2574 $\pm$ 140 &        ...         & -164  \\  
56799.619  &  CAOS   & 1592 $\pm$ 265 & 798    $\pm$ 74  &   -103  & 58267.548  &  CAOS   & 2623 $\pm$ 333 &        ...         & -160  \\  
56816.519  &  CAOS   & 1696 $\pm$ 363 & 701    $\pm$ 30  &   -101  & 58294.013  &  HAR    & 2531 $\pm$ 118 & 520    $\pm$ 15    & -187  \\  
56842.519  &  CAOS   & 1647 $\pm$ 336 & 772    $\pm$ 32  &   -98   & 58332.357  &  CAOS   & 2369 $\pm$ 123 & 733    $\pm$ 33    & -148  \\
\hline
\hline
\end{tabular}
\end{table*}

%% file: Tables/Lines_S_PROCESS_Splitting.tex
\begin{table}
\small
\center
\caption{\label{tab:SProcLines}  Absorption spectral lines of {\it s}-process elements which show important asymmetries and splitting. Wavelengths and involved level energies are from \textsc{nist} database.} 
\begin{tabular}{cccc}
\hline 
\hline
$\lambda_{teo}$ & El & $E_{i}$ & $E_{f}$    \\
$[$\AA$]$      &     & [eV]    & [eV]       \\ 
\hline
4883.682  & YII  & 1.08 & 3.62  \\
4900.118  & YII  & 1.03 & 3.56  \\
5087.418  & YII  & 1.08 & 3.52  \\
5123.209  & YII  & 0.99 & 3.41  \\
4554.033  & BaII & 0.00 & 2.72  \\
5853.675  & BaII & 0.60 & 2.72  \\
6141.713  & BaII & 0.70 & 2.72  \\
6496.898  & BaII & 0.60 & 2.51  \\
\hline
\hline
\end{tabular}
\end{table}

%% file: Tables/table_emission_lines.tex
\begin{table*}
\small
\caption{\label{tab:emission_lines} Emission lines parameters. For each transition the identifications (i.e. theoretical wavelength, element, lower and upper level energy) and the average kinetic parameters (i.e. peak velocity, full-width-at-alf-maximum $\rm FWHM$ and equivalent width $\rm EW$) are reported. $\rm FWHM$ are corrected for the instrumental broadening.}
\begin{tabular}{ccccccc}
\hline
\hline
$\rm \lambda_{teo}$ & El               & $\rm E_{low}$  &  $\rm E_{up}$ & v$\rm _p$           & $\rm FWHM$          & $\rm EW$       \\
$[$\AA$]$           &                  & [eV]         & [eV]         & [$\rm km$ $\rm s^{-1}$]     & [$\rm km$ $\rm s^{-1}$]     & $[$m\AA$]$     \\ 
\hline 
 4102.936           & SiI              & 1.91         & 4.93         & -27.95 $\pm$ 0.12   & 11.19 $\pm$ 1.30   & 48.43  $\pm$ 6.50 \\  
 4976.326           & NiI              & 1.68         & 4.17         & -27.83 $\pm$ 0.25   & 9.35  $\pm$ 1.06   & 9.69   $\pm$ 0.99 \\
 5839.760           & TiI              & 1.46         & 3.58         & -28.36 $\pm$ 0.46   & 9.36  $\pm$ 2.56   & 4.31   $\pm$ 0.84 \\
 5846.994           & NiI              & 1.68         & 3.80         & -28.07 $\pm$ 0.17   & 10.28 $\pm$ 0.78   & 18.10  $\pm$ 1.02 \\
 5866.449           & TiI              & 1.07         & 3.18         & -27.86 $\pm$ 0.16   & 9.13  $\pm$ 0.57   & 19.08  $\pm$ 1.07 \\
 6007.310           & NiI              & 1.68         & 3.74         & -27.34 $\pm$ 0.11   & 8.75  $\pm$ 0.99   & 23.43  $\pm$ 3.08 \\
 6108.120           & NiI              & 1.68         & 3.70         & -28.56 $\pm$ 0.05   & 10.75 $\pm$ 0.66   & 66.26  $\pm$ 3.62 \\
 6572.780           & CaI              & 0.00         & 1.88         & -27.37 $\pm$ 0.12   & 8.79  $\pm$ 0.72   & 32.83  $\pm$ 3.58 \\
 6574.228           & FeI              & 0.99         & 2.87         & -27.88 $\pm$ 0.07   & 9.89  $\pm$ 0.28   & 53.37  $\pm$ 1.56 \\
 6743.119           & TiI              & 0.90         & 2.74         & -27.11 $\pm$ 0.06   & 10.25 $\pm$ 0.48   & 70.35  $\pm$ 3.32 \\
 6771.030           & CoI              & 1.88         & 3.71         & -27.86 $\pm$ 0.23   & 12.06 $\pm$ 1.33   & 25.34  $\pm$ 2.44 \\
 7052.890           & CoI              & 1.96         & 3.71         & -28.93 $\pm$ 0.06   & 12.07 $\pm$ 0.91   & 56.44  $\pm$ 2.52 \\
 7138.903           & TiI              & 1.44         & 3.18         & -27.97 $\pm$ 0.13   & 9.14  $\pm$ 1.18   & 20.56  $\pm$ 1.61 \\
 7216.182           & TiI              & 1.44         & 3.16         & -27.15 $\pm$ 0.18   & 8.27  $\pm$ 0.89   & 18.53  $\pm$ 1.45 \\
 7385.240           & NiI              & 2.74         & 4.42         & -28.67 $\pm$ 0.12   & 10.02 $\pm$ 0.88   & 35.15  $\pm$ 2.08 \\
 7714.320           & NiI              & 1.93         & 3.54         & -28.63 $\pm$ 0.07   & 9.68  $\pm$ 0.60   & 85.90  $\pm$ 4.08 \\
 7788.940           & NiI              & 1.95         & 3.54         & -28.80 $\pm$ 0.09   & 9.94  $\pm$ 0.66   & 70.98  $\pm$ 2.93 \\
 8047.618           & FeI              & 0.86         & 2.39         & -28.04 $\pm$ 0.05   & 10.38 $\pm$ 0.50   & 157.94 $\pm$ 5.89 \\
 8204.102           & FeI              & 0.91         & 2.42         & -27.67 $\pm$ 0.13   & 8.82  $\pm$ 0.76   & 38.22  $\pm$ 2.65 \\
 8364.237           & TiI              & 0.83         & 2.31         & -27.50 $\pm$ 0.14   & 6.96  $\pm$ 0.59   & 27.75  $\pm$ 2.08 \\
 8365.633           & FeI              & 3.25         & 4.73         & -28.49 $\pm$ 0.36   & 8.46  $\pm$ 1.26   & 14.11  $\pm$ 1.96 \\
 8412.356           & TiI              & 0.81         & 2.29         & -27.71 $\pm$ 0.11   & 8.34  $\pm$ 0.56   & 48.80  $\pm$ 2.63 \\
 8434.959           & TiI              & 0.85         & 2.32         & -27.76 $\pm$ 0.28   & 8.28  $\pm$ 0.69   & 53.84  $\pm$ 6.28 \\
 8435.650           & TiI              & 0.84         & 2.30         & -27.62 $\pm$ 0.07   & 8.30  $\pm$ 0.45   & 56.92  $\pm$ 4.79 \\
 8442.970           & TiI              & 2.16         & 3.87         & -27.33 $\pm$ 0.20   & 7.53  $\pm$ 0.86   & 19.82  $\pm$ 1.92 \\
 8450.890           & TiI              & 2.25         & 3.71         & -28.73 $\pm$ 0.15   & 8.31  $\pm$ 0.69   & 32.12  $\pm$ 2.75 \\
 8494.448           & TiI              & 1.74         & 3.20         & -28.06 $\pm$ 0.19   & 11.24 $\pm$ 0.97   & 31.54  $\pm$ 3.96 \\
 8518.352           & TiI              & 1.88         & 3.33         & -28.07 $\pm$ 0.10   & 9.75  $\pm$ 0.79   & 78.44  $\pm$ 3.04 \\
 8548.088           & TiI              & 1.87         & 3.32         & -27.77 $\pm$ 0.07   & 9.53  $\pm$ 0.65   & 90.55  $\pm$ 4.35 \\
 8565.488           & TiI              & 1.74         & 3.19         & -27.25 $\pm$ 0.28   & 7.35  $\pm$ 1.21   & 12.46  $\pm$ 1.78 \\
 8757.187           & FeI              & 2.84         & 4.26         & -28.38 $\pm$ 0.10   & 9.08  $\pm$ 0.96   & 36.48  $\pm$ 2.42 \\
 9675.547           & TiI              & 0.84         & 2.12         & -27.46 $\pm$ 0.15   & 9.08  $\pm$ 0.71   & 69.18  $\pm$ 4.79 \\
\hline
\hline
\end{tabular}
\end{table*}